\documentclass[10pt,prx,twocolumn,superscriptaddress,aps]{revtex4-2}

\usepackage{graphicx}
\usepackage{dcolumn}
\usepackage{bm}
\usepackage{amsmath}
\usepackage{amsfonts}
\usepackage{amssymb}
\usepackage{dcolumn}
\usepackage{epsfig}
\usepackage{graphicx}
\usepackage{latexsym}
\usepackage{setspace}
\usepackage{fancyhdr}
\usepackage{wrapfig}
\usepackage{textgreek}
\usepackage{ulem}
\usepackage{titlesec}
\usepackage{outlines}
\usepackage{multirow}
\usepackage{adjustbox}
\usepackage{enumitem}
\usepackage[official]{eurosym}
\usepackage{mdframed}
\usepackage{enumitem}
\usepackage{braket}
\usepackage{hhline}
\usepackage[title]{appendix}
\usepackage{upgreek}
\usepackage{comment}
\usepackage{titlesec}
\titlespacing{\section}{0pt}{1.5em}{1.5em}
\titlespacing{\subsection}{0pt}{1.5em}{1.5em}

\usepackage{hyperref}
\hypersetup{
    colorlinks=true,
    citecolor=blue,
    linkcolor=blue,
    filecolor=blue,      
    urlcolor=black,
    linktoc=all
}

\begin{document}

\pagestyle{plain}
\title{Singly resonant InGaP microresonators for efficient, low-threshold second-order nonlinear optics}
\author{Yiming Pang}
\email{yimingpang@ucsb.edu}
\affiliation{\protect\hbox{Electrical and Computer Engineering Department, University of California, Santa Barbara, CA 93106, USA}}
\author{Xuefeng Li}
\affiliation{\protect\hbox{Electrical and Computer Engineering Department, University of California, Santa Barbara, CA 93106, USA}}
\author{Lucas Wang}
\affiliation{Physics Department, University of California, Santa Barbara, CA 93106, USA}
\author{Lillian Thiel}
\affiliation{\protect\hbox{Electrical and Computer Engineering Department, University of California, Santa Barbara, CA 93106, USA}}
\author{Melissa A. Guidry}
\affiliation{\protect\hbox{LIGO, Massachusetts Institute of Technology, Cambridge, Massachusetts, USA}}
\author{Joshua E. Castro}
\affiliation{\protect\hbox{Electrical and Computer Engineering Department, University of California, Santa Barbara, CA 93106, USA}}
\author{\\Max Meunier}
\affiliation{\protect\hbox{Electrical and Computer Engineering Department, University of California, Santa Barbara, CA 93106, USA}}
\author{Nergis Mavalvala}
\affiliation{\protect\hbox{LIGO, Massachusetts Institute of Technology, Cambridge, Massachusetts, USA}}
\author{John E. Bowers}
\affiliation{\protect\hbox{Electrical and Computer Engineering Department, University of California, Santa Barbara, CA 93106, USA}}
\affiliation{Materials Department, University of California, Santa Barbara, CA 93106, USA}
\author{Galan Moody}
\email{moody@ucsb.edu}
\affiliation{\protect\hbox{Electrical and Computer Engineering Department, University of California, Santa Barbara, CA 93106, USA}}

\begin{abstract}
\noindent Strong $\chi^{(2)}$ interactions at low optical powers are key to scalable nonlinear and quantum photonics. Although doubly resonant microresonators exhibit exceptional efficiencies, they require simultaneous resonance and phase matching of widely separated optical frequencies, making them inherently sensitive to fabrication and operating conditions. Singly resonant cavities eliminate this constraint by resonating at only the fundamental frequency while the second-harmonic field propagates in a single pass, but they have traditionally sacrificed nonlinear efficiency. Here, we demonstrate a singly resonant InGaP-on-insulator microresonator that overcomes this tradeoff. Combining the large $\chi^{(2)}$ nonlinearity of InGaP with high-quality-factor resonators, we achieve efficient and widely tunable second-harmonic generation, broadband quantum-frequency-comb generation, and low-threshold optical parametric oscillation. Unlike conventional centimeter-long singly resonant devices, our millimeter-scale InGaP devices deliver nonlinear performance comparable to, and even exceeding, doubly resonant systems. By overcoming the conventional efficiency-robustness tradeoff of singly resonant devices, this work opens a new route to low-power, broadband, and widely tunable $\chi^{(2)}$ nonlinear and quantum photonics.

\end{abstract}
\maketitle

\thispagestyle{plain}


\section{Introduction}
Cavity-enhanced second-order nonlinear ($\chi^{(2)}$) interactions underpin various optical phenomena in both the classical and quantum regimes, including frequency conversion between disparate spectral ranges, generation of parametric oscillation and broadband frequency combs, and quantum light sources for discrete- and continuous-variable entanglement distribution \cite{moody20222022, dutt2024nonlinear}. Over the past decade, doubly resonant microresonators have become widely adopted for efficient second-order nonlinear optical processes. Resonating both the fundamental-frequency (FF) and the second-harmonic (SH) fields, these devices deliver strong coupling between the FF and SH modes and high nonlinear efficiencies. Such doubly resonant microresonators have been demonstrated across numerous integrated material platforms, including AlN \cite{guo2016second, bruch201817}, LiNbO$_{3}$ \cite{lu2020toward, lu2019periodically, chen2021efficient, wu2024second, du2024high}, InGaP \cite{li2026high, akin2024ingap, zhao2022ingap}, Si$_{3}$N$_{4}$ \cite{lu2021efficient, nitiss2022optically, yuan2025efficient, clementi2025ultrabroadband}, GaP \cite{logan2018400}, GaAs \cite{chang2019strong} and SiC \cite{lukin20204h}, as well as the emerging LiTaO$_{3}$ \cite{mohanraj2026integrated} and AlScN \cite{liu2026hybrid} platforms. Among these materials, InGaP combines an exceptionally large second-order nonlinear susceptibility ($\chi^{(2)}=220$~pm/V) with a high refractive index ($n > 3$) across a wide transparency window (0.65~$\upmu$m to 11~$\upmu$m) \cite{akin2024perspectives}, standing out as an attractive choice for realizing compact, efficient nonlinear photonic devices \cite{akin2024ingap, li2026high, ahler2026low}. 

Doubly resonant operation, however, imposes several practical constraints. Spectrally, a set of phase-matched FF and SH resonances must be tuned into simultaneous alignment, which, due to fabrication imperfections, often requires differential tuning of the two modes with disparate thermo-optic responses, making this alignment sensitive to temperature and optical power. When operating for second-harmonic generation (SHG), the large intracavity SH field leads to depletion of the FF pump field \cite{guo2016second}. Since the onset scales with conversion efficiency, the most power-efficient microrings deplete at the lowest pump power, limiting their maximum SH output \cite{lu2020toward}.

\begin{figure*}[t!]
    \centering
    \includegraphics[width=\textwidth]{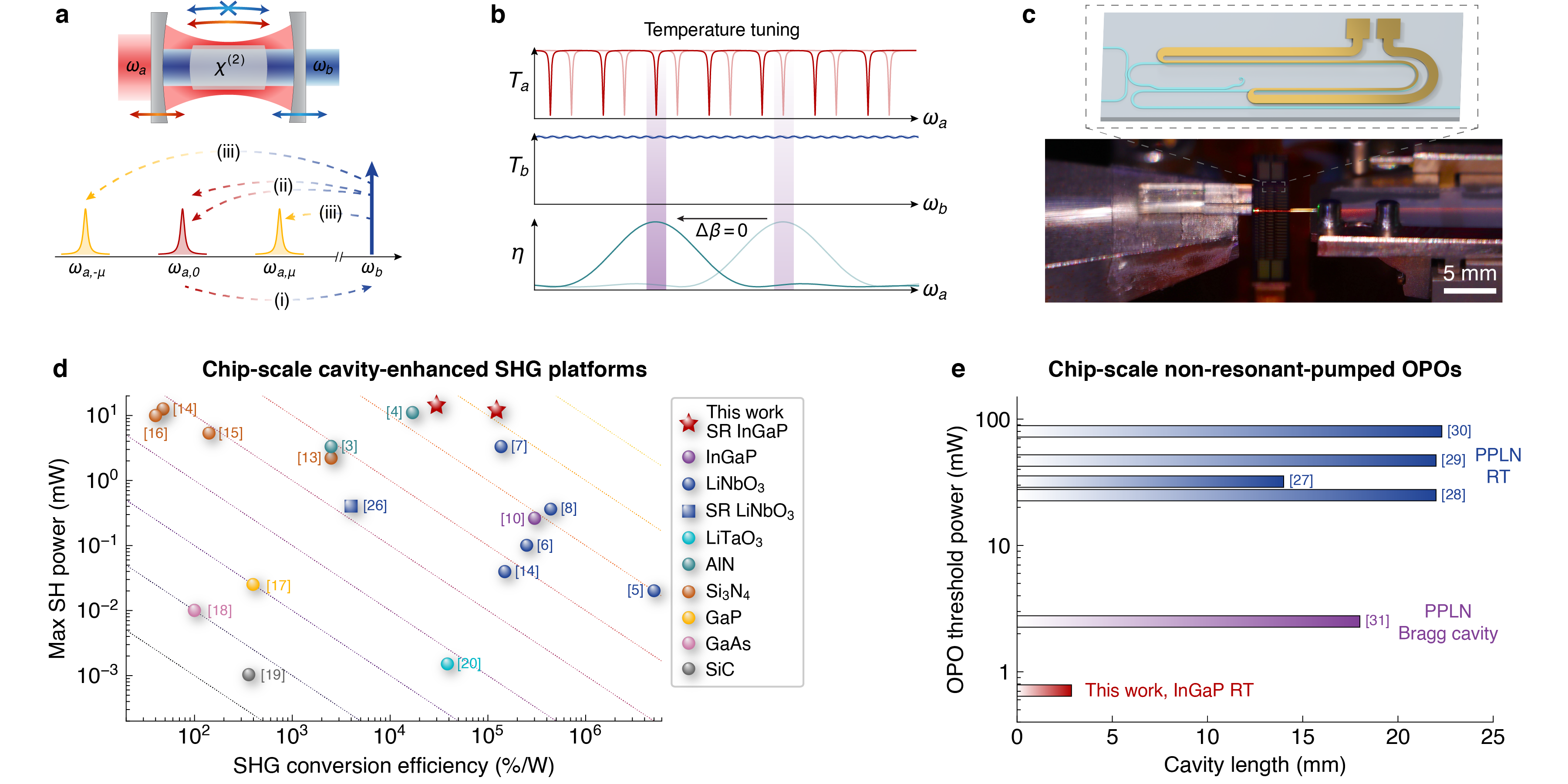}
    \vspace{-15pt}
    \caption{\small \label{Fig1} \textbf{Chip-scale singly resonant $\chi^{(2)}$ nonlinear optical cavity.}
    \textbf{a}, The schematics of a singly resonant $\chi^{(2)}$ cavity consisting of a $\chi^{(2)}$ material embedded in two wavelength-selective mirrors, and the schematics of nonlinear interactions including: (i) second-harmonic generation (SHG); (ii) degenerate, and (iii) non-degenerate parametric down-conversion (SPDC) or optical parametric oscillation (OPO), depending on the pump power level. Cascaded SPDC/OPO can take place through back-to-back SHG and non-degenerate SPDC/OPO. \textbf{b}, Schematic illustration of transmission spectra showing resonances of the FF modes ($T_{a}$) and non-resonant behavior of the SH mode ($T_{b}$). Maximum nonlinear coupling ($\eta$) is achieved when there is zero phase mismatch $\Delta \beta=0$ among the interacting waves. Temperature tuning of the cavity shifts both FF resonances and the phase-matching condition. \textbf{c}, An image of a 4~mm $\times$ 20~mm InGaP-on-insulator (InGaPOI) photonic chip with more than 120 microresonators of the type illustrated on the top panel. Bright, near-visible red light is generated from the resonator when pumped at telecom wavelength. \textbf{d}, Comparison of SHG conversion efficiency and maximum on-chip SHG power for chip-scale cavity-enhanced SHG platforms, including both doubly resonant microresonators \cite{guo2016second, bruch201817, lu2020toward, lu2019periodically, chen2021efficient, wu2024second, du2024high, li2026high, logan2018400, chang2019strong, lu2021efficient, nitiss2022optically, yuan2025efficient, clementi2025ultrabroadband, lukin20204h, mohanraj2026integrated} and singly resonant (SR) devices presented in this work and in \cite{hefti2026enhancing}. Dotted lines are contours of constant product of the two axes spaced by an order of magnitude. \textbf{e}, Comparison of cavity length and OPO threshold power for chip-scale non-resonant-pumped OPO platforms \cite{ledezma2023octave, park2024single, stokowski2024integrated, hwang2023mid, kellner2025low}. PPLN, periodically-poled lithium niobate; RT, racetrack microresonator.
    }
    \vspace{-10pt}
\end{figure*}

Alternatively, these limitations are relaxed in a \textit{singly resonant cavity} that enhances only the FF fields while the SH field remains non-resonant via a wavelength-selective dichroic mirror as shown in Fig.~\ref{Fig1}\textbf{a},\textbf{b} \cite{kozlovsky1988efficient, collett1991two}. This configuration enables independent adjustment of the optimal phase-matching condition without additionally needing to satisfy the doubly resonant condition, an approach recently demonstrated on chip in periodically poled lithium niobate (PPLN) with broadband SHG~\cite{hefti2026enhancing}.

Pumped instead at the SH, the same cavity operates as an optical parametric oscillator (OPO) whose signal and idler modes are resonant while the pump is non-resonant. This architecture has enabled tunable and broadband classical \cite{ledezma2023octave,stokowski2024integrated, hwang2023mid, kellner2025low} and quantum light sources \cite{park2024single}, with prospects for extending coherent light generation to wavelengths that are challenging for conventional semiconductor laser gain media \cite{hwang2023mid}. Despite these advances, previous demonstrations have largely relied on PPLN with centimeter-long cavities, exhibiting typical threshold powers of tens of milliwatts \cite{ledezma2023octave, park2024single, stokowski2024integrated, hwang2023mid} -- two-to-three orders of magnitude higher than those of state-of-the-art doubly resonant microrings \cite{bruch2019chip}. 

In this work, we narrow the performance gap between singly and doubly resonant architectures by demonstrating millimeter-scale, singly resonant InGaP microresonators for efficient SHG, multimode photon-pair generation, and low-threshold OPO. As shown in Fig.~\ref{Fig1}\textbf{c}, the singly resonant device combines modal-phase-matched nonlinear waveguide sections with a dichroic coupler in a racetrack resonator to resonate the FF fields selectively while remaining non-resonant at the SH frequencies, reaching intrinsic quality factors up to $1.2\times10^{6}$ at FF. When pumped at the telecom wavelengths, a high SHG conversion efficiency of $1.23\times10^{5}$~\%/W is achieved in the non-depletion regime, while the SHG wavelength is tunable across 1.8~THz near 780~nm. Above 25~mW of pump power, full extraction of the SH output from the resonator allows $>10$~mW of SHG power. Figure~\ref{Fig1}\textbf{d} places our device performance among other chip-scale cavity-enhanced SHG platforms, showing that our singly resonant devices match the conversion efficiency of doubly resonant systems while exceeding their SH output power.

Under SH pumping, the resonator produces quantum frequency combs (QFCs) via spontaneous parametric down-conversion (SPDC) with more than 25 sets of time-energy-correlated photon pairs, with an on-chip pair generation efficiency of up to 0.26~MHz/$\upmu$W per mode pair. The QFC profile can be spectrally controlled by GHz-level detuning of the SH pump frequency. At increased SH pump power, a compact 2.86~mm long resonator generates parametric oscillation with an OPO threshold of 0.71~mW. As summarized in Fig.~\ref{Fig1}\textbf{e}, this is to our knowledge the lowest threshold reported for any chip-scale OPO with a non-resonant pump, and notably, achieved in a cavity at least five times shorter than those previously employed in PPLN photonics. Depending on SH pump detuning, the degenerate oscillation transitions into non-degenerate operation with a maximum signal-idler separation of 5.7~THz, and frequency comb formation is observed at optimal detuning. We further model these nonlinear phenomena using coupled-mode theory, showing good agreement between the model and experiment. These results establish singly resonant InGaP microresonators as a versatile platform for integrated second-order nonlinear optics.

\begin{figure*}[t!]
    \centering
    \includegraphics[width=\textwidth]{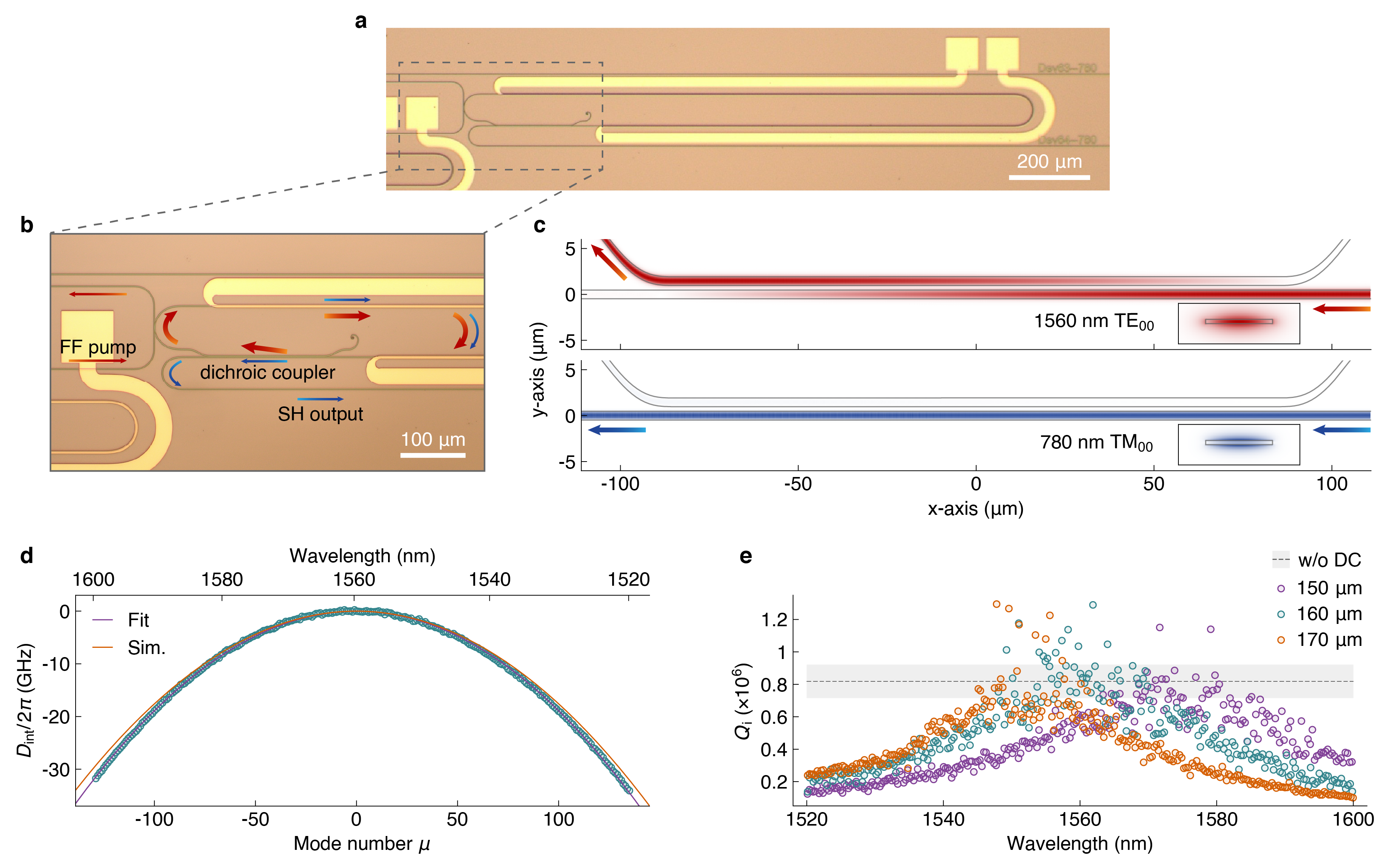}
    \vspace{-15pt}
    \caption{\small \label{Fig2} \textbf{Device design and linear characterization.} \textbf{a}, Microscope image of a singly resonant resonator with cavity length of 2.86~mm. \textbf{b}, Zoomed-in view of the coupling section when the device operates in an SHG process, where red and blue arrows show the direction of light propagation for the FF pump and the SH output, respectively. The FF and SH fields are separated at the dichroic coupler (DC), leading to resonant enhancement of the FF mode while the SH mode is non-resonant and extracted out of the resonator. \textbf{c}, Simulated field intensity at the DC when the input field is 1560~nm TE$_{00}$ mode (top) and 780~nm TM$_{00}$ mode (bottom). Insets show cross-sectional mode profiles of the respective optical mode in a straight waveguide. \textbf{d}, Measured and simulated integrated dispersion of a 37~GHz resonator showing normal dispersion of the FF mode. \textbf{e}, Intrinsic quality factor of resonators with DC lengths of 150~$\upmu$m, 160~$\upmu$m, and 170~$\upmu$m. The gray dashed line and gray shaded region indicate the mean and one standard deviation of $Q_{\mathrm{i}}$ from an all-pass racetrack resonator without the DC.}
    \vspace{-10pt}
\end{figure*}
\section{Results}
\subsection{Device design}
\label{sec:design}
The singly resonant devices are fabricated using the InGaP-on-insulator (InGaPOI) photonic platform, with details on the fabrication processes provided in Section~\ref{sec:fabrication}. Figure~\ref{Fig2}\textbf{a} and~\ref{Fig2}\textbf{b} show microscope images of a fabricated device with a cavity length ($L_{\mathrm{c}}$) of 2.86~mm, consisting of a pulley-coupled racetrack resonator with a directional coupler embedded in one of the straight sections. To utilize the $\chi^{(2)}_{xyz}$ component of the susceptibility tensor, we design the waveguide dimensions to achieve the modal phase-matching condition for the 1560~nm TE$_{00}$ and 780~nm TM$_{00}$ modes by setting the phase-mismatch to zero, $\Delta \beta=\beta_{b}-2\beta_{a}=0$, where $\beta_{a}$ and $\beta_{b}$ are the phase constants of the FF and SH modes, respectively. In this case, the InGaP thickness is 105~nm and a waveguide width of 1100~nm is chosen for the phase-matched straight sections within the resonator. A waveguide width of 950~nm is utilized for the coupling sections at the pulley coupler and the dichroic coupler (DC), with an adiabatic taper placed between the nonlinear and linear sections. The DC is optimized such that when the input optical mode in the bottom waveguide is the 1560~nm TE$_{00}$ mode, the field crosses over to the top waveguide; in contrast, when the input is the 780~nm TM$_{00}$ mode, the field remains in the bottom waveguide. Coupling of the two optical modes at the DC is simulated using finite-difference time-domain (FDTD) methods, as shown in Fig.~\ref{Fig2}\textbf{c}. The simulated cross-coupling coefficient for the 1560~nm TE$_{00}$ mode is $k^{2}_{\text{dc}, a}=99.7\%$, while the self-coupling coefficient for the 780~nm TM$_{00}$ mode is $r^{2}_{\text{dc}, b}=98.8\%$. The pulley coupler bus waveguide has a width of 770~nm with varying gaps and coupling angles to realize devices with the ideal coupling condition for various experiments studied in this work: under-coupling for low-threshold OPO; critical-coupling for efficient SHG in the non-depletion regime; and over-coupling for raising the maximum SH output power.
\subsection{Linear characterization}
\label{sec:linear}
Linear characterization of the fabricated device is performed by measuring its transmission spectra from 1520~nm to 1600~nm with a tunable continuous-wave (CW) laser. The resonant frequencies for FF modes near a central frequency $\omega_{a,0}$ can be written as
\begin{equation}
    \omega_{a,\mu}=\omega_{a,0}+D_{1}\mu+\frac{D_{2}}{2}\mu^{2}+\dots, \quad \mu\in \mathbb{Z},
\end{equation}
where $\mu$ is the relative mode number, $D_{1}/2\pi$ is the free-spectral range (FSR), and $D_{2}$ is the second-order dispersion parameter. The deviation of FF resonances from an equidistant frequency grid is known as the integrated dispersion, defined as $D_{\mathrm{int}}=\omega_{a,\mu}-(\omega_{a,0}+D_{1}\mu)$. The measured and simulated integrated dispersion of a 2.86~mm long racetrack resonator is shown in Fig.~\ref{Fig2}\textbf{d}, where an FSR of 37.22~GHz is obtained and a quadratic fit yields normal dispersion with $D_{2}/2\pi=-3.75$~MHz. Subtle deviations of the measured dispersion from the fit originate from roughness-induced backscattering or coupling to higher-order modes.

By fitting the resonance lineshapes (see Supplementary Information Section S3), the intrinsic linewidth ($\kappa_{a,\mathrm{i}}$) and extrinsic linewidth ($\kappa_{a,\mathrm{e}}$) of the FF modes are extracted. In  Fig.~\ref{Fig2}\textbf{e}, we compare the intrinsic quality factors $Q_{a,\mathrm{i}}=\omega_{a}/\kappa_{a,\mathrm{i}}$ from three singly resonant devices with different DC lengths (150~$\upmu$m, 160~$\upmu$m, and 170~$\upmu$m). We observe a wavelength-dependent $Q_{a,\mathrm{i}}$ that is due to varying coupling ratio of the dichroic coupler: $k^{2}_{\text{dc}, a} = \sin^{2}(\pi L_{\mathrm{dc}}\Delta n_{a}/\lambda_{a})$, where $\Delta n_{a}$ is the effective index difference of the supermodes and it increases with wavelength. Therefore, for a fixed coupler length, longer wavelength modes have stronger coupling, and $Q_{a,\mathrm{i}}$ is maximized when $k^{2}_{\text{dc}, a} \rightarrow 1$. For each design, there is a $\sim20$~nm window where $Q_{a,\mathrm{i}}$ exceeds $6\times10^{5}$, with some resonances showing $Q_{a,\mathrm{i}}$ above $1.2\times10^{6}$. As indicated by the gray dashed line, a conventional all-pass racetrack resonator of the same cavity length but without the DC shows a mean $Q_{a,\mathrm{i}}$ of $8\times10^{5}$, which is comparable those of the singly resonant devices at optimized wavelength. Hence, we conclude that the DC introduces insignificant insertion loss other than imposing the wavelength-dependent coupling rate. The wavelength dependence could be reduced using adiabatic couplers with broadband coupling \cite{guo2017parametric}, albeit at the cost of a longer coupler length that increases device footprint. These quality factors are among the highest values reported in InGaP integrated photonics to date \cite{li2026high, akin2024ingap, chopin2023ultra}, which, when combined with the large $\chi^{(2)}$ nonlinearity, enhances the device's nonlinear performance. 

\subsection{Second-harmonic generation}
\label{sec:SHG}
\begin{figure*}[t!]
    \centering
    \includegraphics[width=\textwidth]{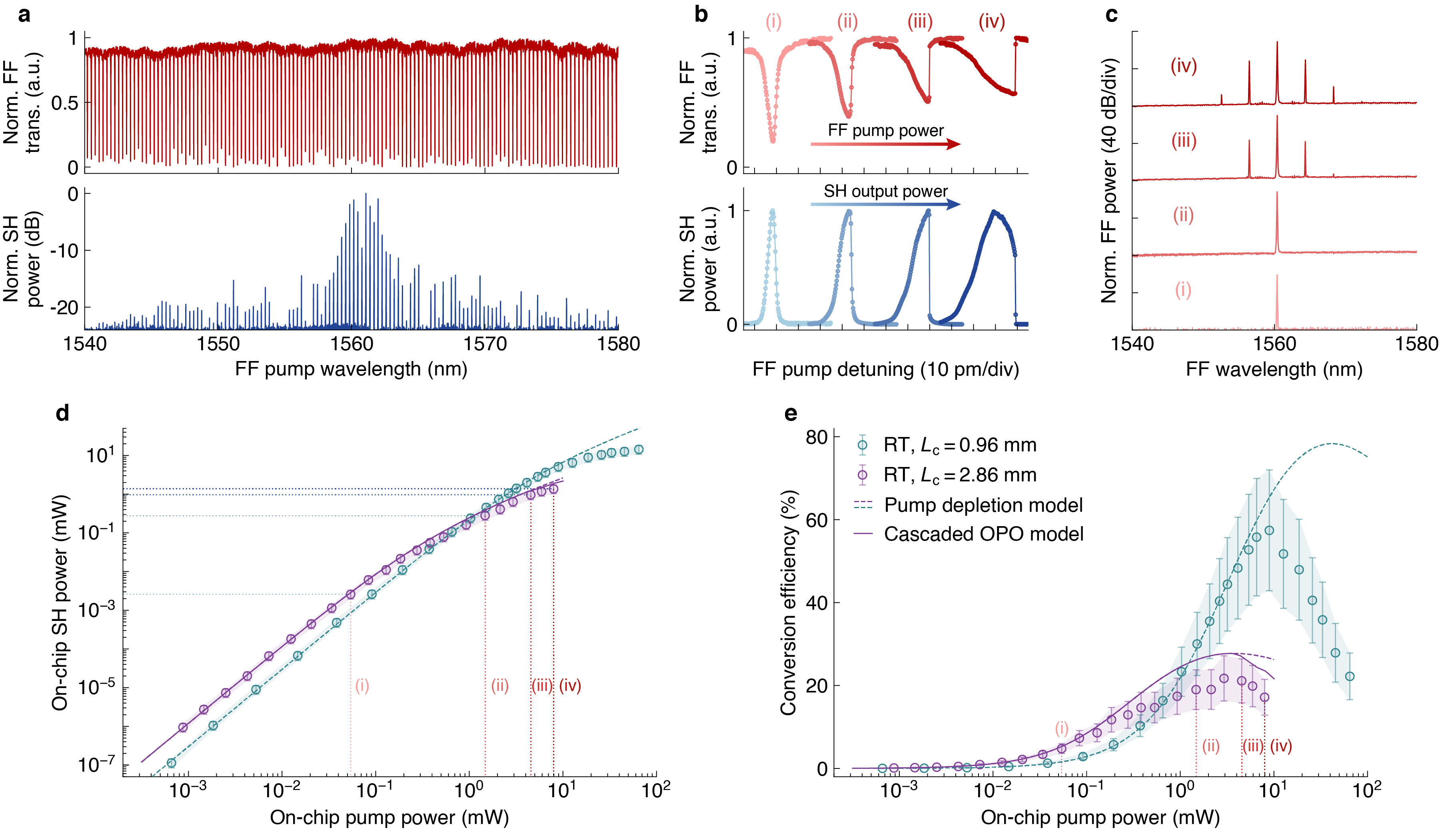}
    \vspace{-15pt}
    \caption{\small \label{Fig3} \textbf{Second-harmonic generation.} \textbf{a}, Normalized transmission of the FF pump (top) and normalized SH output power (bottom) versus pump wavelength for a critically coupled 37~GHz resonator. \textbf{b}, Pump transmission and SH lineshape as the laser is tuned from lower to higher wavelength near the FF resonance at various pump power levels (i)-(iv). \textbf{c}, Power spectra at the FF output taken before the laser falls off the thermally pulled pump resonance. At low pump power, only the transmitted pump near 1560~nm is present, see spectrum (i) and (ii). At pump power greater than 4.53~mW, cascaded optical parametric oscillation at non-degenerate FF modes is observed, as shown in spectrum (iii) and (iv). \textbf{d}, On-chip SH output power, and \textbf{e}, conversion efficiency versus the on-chip pump power for a racetrack resonator with cavity length ($L_{\mathrm{c}}$) of an overcoupled 0.96~mm (green) and an undercoupled 2.86~mm long racetrack resonator (purple). Dashed curves are fits using the pump depletion model, and the solid curve uses the model that include cascaded OPO. Dotted lines in \textbf{d} and \textbf{e} indicate the FF pump and SH output power levels (i)-(iv) in \textbf{b} and \textbf{c}.
    }
    \vspace{-10pt}
\end{figure*}
To characterize the nonlinear performance, we pump the device at the FF resonances with a CW laser and collect the generated SH light at the output. Figure~\ref{Fig3}\textbf{a} shows an example transmission spectrum of the FF pump and the output SH spectra as the pump is swept from 1540~nm to 1580~nm. The SH output wavelength can be tuned freely across 1.8~THz from 779.8~nm to 783.5~nm by sourcing a direct current on the integrated thermo-optic heaters on top of the resonator while the chip temperature is fixed at 20~$^{\circ}$C, see Supplementary Information Section S4. 

Power-dependent measurements are performed by controlling the incident pump power with a variable optical attenuator (VOA), and the results are summarized in Fig.~\ref{Fig3}\textbf{b}-\textbf{e}. Here, we focus our discussion on an undercoupled, 2.86-mm-long resonator with FSR of 37~GHz that is later studied in the SPDC and OPO experiments. Figure~\ref{Fig3}\textbf{b} shows normalized transmission spectra of the FF pump and the generated SH lineshapes at various pump power as the laser is swept across one resonance. As pump power increases, the FF lineshapes deviate from a perfect Lorentzian and become thermally pulled due to absorption-induced cavity heating and self-phase modulation, and at larger laser detuning, the pump falls off resonance. The SH lineshapes, however, initially follow the FF lineshapes, but deviations are observed for (iii) and (iv) when pump power exceeds 4.53~mW, where the SH power drops before the pump falls off the FF resonance. Power spectra of the transmitted FF pump before it falls off resonance are recorded on an optical spectrum analyzer (OSA), as shown in Fig.~\ref{Fig3}\textbf{c}. At low pump power (i, ii), only the pump near 1560~nm is resolved, and at higher power (iii, iv), symmetric sidebands emerge at mode pair $\mu=\pm13$ and $\mu=\pm26$ due to cascaded OPO \cite{schiller1996subharmonic}. At high FF pump power, the generated SH introduces instability of non-degenerate mode pairs at $\omega_{a,\pm\mu}$, producing parametric oscillations that deplete the SH field and distort its lineshape. Thermal self-stability pins the laser frequency to the blue side of the thermally shifted resonance, which disfavors energy matching for normal dispersion (see Section \ref{sec:OPO}). Nevertheless, oscillation occurs because the resonant frequencies of individual modes deviate from the fitted dispersion, an effect also observed in doubly resonant microresonators \cite{mckenna2022ultra}.

After accounting for coupling losses and setup losses, we calculate the on-chip pump power and on-chip SH output power, shown in Fig.~\ref{Fig3}\textbf{d}, along with the absolute conversion efficiency versus pump power in Fig.~\ref{Fig3}\textbf{e}. The SH output power scales with pump power quadratically in the non-depletion regime, with an SHG conversion efficiency of (see Supplementary Information Section S1):
\begin{equation}
\eta_{\mathrm{SHG}}=\frac{P_{b, \mathrm{out}}}{P_{a, \mathrm{in}}^{2}}=\eta_{0}L^{2}|\Gamma|^{2}\frac{16\kappa_{a, \mathrm{e}}^{2}}{\kappa_{a}^{4}\tau_{a}^{2}}.
\end{equation}
where $\eta_{0}$ is the normalized waveguide SHG efficiency in units of W$^{-1}$cm$^{-2}$, $L$ is the length of phase-matched nonlinear section, $\kappa_{a}=\omega_{a}/Q_{a}$ is the total loss rate, and $\tau_{a}$ is the round-trip time. We introduce $\Gamma$ as a dimensionless nonlinear coupling ideality factor that accounts for SH propagation loss $\alpha_{b}$ and phase-mismatch $\Delta\beta$
\begin{equation}
    \Gamma=\frac{1-e^{-(\alpha_{b}/2+i\Delta \beta)L}}{(\alpha_{b}/2+i\Delta \beta)L }.
\end{equation}

For the undercoupled 2.86~mm long resonator with cavity escape efficiency of $\eta_{a,\mathrm{esc}}=\kappa_{a,\mathrm{e}}/\kappa_{a}=0.32$ and $Q_{a,\mathrm{i}}=6.29\times10^{5}$, we measured an SHG conversion efficiency of $(1.20\pm0.12)\times10^{5}$~\%/W, and a maximum conversion efficiency of $\approx22$\% is achieved at a pump power of 2.93~mW. Using measured SH propagation loss of around 8.7~dB/cm and assuming perfect phase-matching, this corresponds to a normalized efficiency of $\eta_{0}=1.0\times10^{4}$~\%W$^{-1}$cm$^{-2}$ in the nonlinear section, which is lower than the theoretical value of $3.1\times10^{4}$~\%W$^{-1}$cm$^{-2}$ based on mode overlap simulations. The discrepancy is likely from small variations of InGaP film thickness along the propagation that result in accumulation of phase-mismatch and reduces the overall nonlinear efficiency, and can be further improved by adaptive fabrication \cite{chen2024adapted} and adaptive tuning \cite{hu2026efficient} techniques. With the extracted $\eta_{0}$, the measured data is overlaid with our theory models (see Supplementary Information Section S1) in Fig.~\ref{Fig3}\textbf{d},\textbf{e}, one considering only pump depletion (dashed lines) while the other includes cascaded OPO that depletes the SHG (solid line). The cascaded OPO threshold determined from the model is about 3.7~mW of FF pump power, in reasonable agreement with our experimental observation at 4.53~mW of input power. 

While the maximum SHG power for the undercoupled device is about 1.4~mW, limited by the onset of cascaded OPO, this can be improved by increasing escape efficiency of the FF mode. For a critically coupled 2.86~mm long device with $Q_{a,\mathrm{i}}=8.12\times10^{5}$, an on-chip SH output power of 12.1~mW is generated with pump power of 57.7~mW, while the same device exhibits high conversion efficiency of $(1.23\pm0.15)\times10^{5}$~\%/W in the non-depletion regime (see Supplementary Information Section S4). Higher SH output power and absolute conversion efficiency can also be achieved in resonators with smaller cavity length, which mitigates nonlinear loss from SH back-conversion (an effective two-photon loss channel) and cascaded OPO formation. As shown by the green data points in Fig.~\ref{Fig3}\textbf{d},\textbf{e}, a 0.96~mm long resonator reaches a maximum absolute conversion efficiency of 57\% and maximum SH output power of 14.4~mW. While no cascaded OPO is observed, the sudden drop in efficiency is potentially due to competing nonlinear effects driven by the large field-enhancement of the pump that leads to parasitic loss of the SH mode \cite{nie2025dissipative}.

\subsection{Quantum frequency comb}
\label{sec:QFC}
\begin{figure*}[t!]
    \centering
    \includegraphics[width=\textwidth]{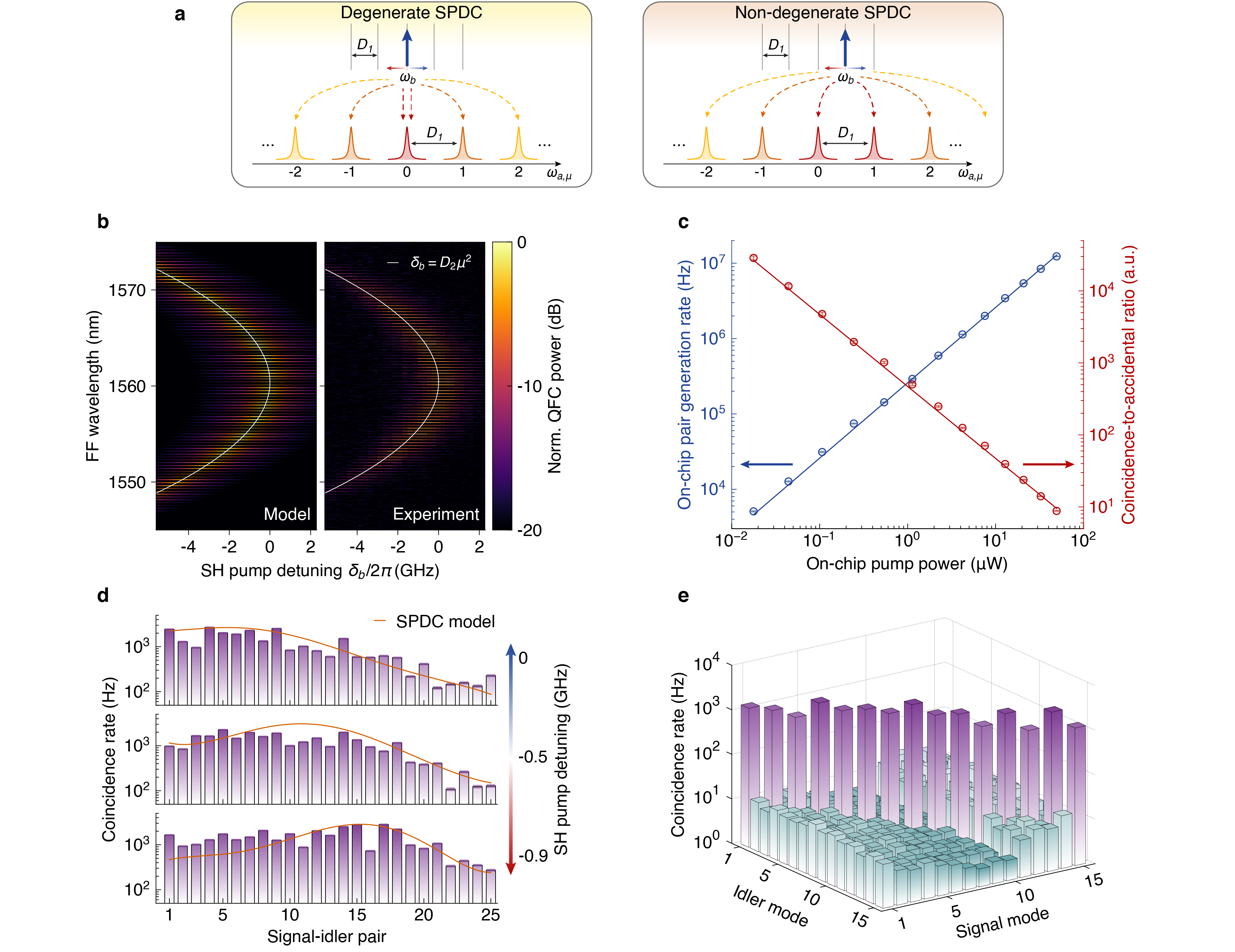}
    \vspace{-15pt}
    \caption{\small \label{Fig4} \textbf{Below-threshold quantum frequency comb and photon-pair generation.} \textbf{a}, Schematic of the two configurations for parametric down-conversion. For a SH pump at an even grid index $m$, the device undergoes degenerate spontaneous parametric down-conversion (SPDC) at $\omega_{a,0}$ and non-degenerate SPDC at signal-idler pair ($-\mu$,~$\mu$); for a pump at odd $m$ it undergoes non-degenerate SPDC only ($m=1$ case is illustrated) at the pump-symmetric mode pairs ($m-\mu$,~$\mu$). \textbf{b}, Model (left) and experimental measurement (right) of the quantum-frequency-comb power versus SH pump-frequency detuning $\delta_{b}/2\pi$ for the 37~GHz resonator with the degenerate mode near 1560.4~nm. The white curve marks the wavelength of mode pairs satisfying $\delta_{b}=D_{2}\mu^{2}$. \textbf{c}, Measured on-chip pair-generation rate (blue) and coincidence-to-accidental ratio (red) versus on-chip pump power for a set of non-degenerate modes. An on-chip pair-generation efficiency is measured to be 0.26~MHz/$\upmu$W. \textbf{d}, Coincidence rate versus signal–idler mode-pair indices 1–25 at three different pump detuning; orange curve is calculation using the SPDC model. \textbf{e}, Joint spectral correlation of signal and idler modes 1 to 15 at optimized pump detuning, showing bright coincidences along the anti-diagonal of the $15\times15$ frequency space.
    }
    \vspace{-10pt}
\end{figure*}
We now use the device for SPDC by injecting a SH pump and collecting at the FF output. Far below the OPO threshold, a QFC emerges with dense comb lines consisting of time-energy-correlated photon pairs seeded from vacuum fluctuations in discrete spectral modes \cite{caspani2016multifrequency}. Depending on the pump frequency $\omega_{b}$, the device generates QFCs with or without the degenerate mode, as illustrated in Fig.~\ref{Fig4}\textbf{a}. 

We derive the SPDC photon generation rate (see Supplementary Information Section S1) by first defining an equidistant pump frequency grid centered around $2\omega_{a,0}$ for the SH pump,
\begin{equation}
    \omega_{b}^{(m)}=2\omega_{a,0}+D_{1}m, \qquad m\in \mathbb{Z}.
\end{equation}
Cavity-enhanced SPDC occurs only if the pump frequency is tuned near $\omega_{b}^{(m)}$ due to energy conservation. We define a global pump detuning $\delta_{b}=\omega_{b}-2\omega_{a,0}$, and when the pump frequency is near the degenerate mode ($m=0$), the modal detuning and phase-mismatch due to dispersion for the pair of modes at $\omega_{a,\pm\mu}$ labeled as (-$\mu$,~$\mu$) is
\begin{align}
\begin{split}
\delta_{\mu} &=\frac{D_{2}}{2}\mu^{2}-\frac{\delta_{b}}{2},\\
\Delta \beta_{\mu}&\approx\beta_{b}^{(1)}\delta_{b}+\frac{\beta_{b}^{(2)}}{2}\delta_{b}^{2}+\frac{2\pi D_{2}}{L_{\mathrm{c}}D_{1}}\mu^{2},
\end{split}
\end{align}
where $\beta_{b}^{(n)}=\partial^{n}\beta_{b}/\partial\omega^{n}$ is the dispersion parameter at the second harmonic. Assuming an undepleted SH pump field $A_{b,0}=\sqrt{P_{b, \mathrm{in}}}e^{i\phi_{b}}$, the single-pass parametric gain exerted on the signal-idler pair is
\begin{equation}
G_{\mu}=\sqrt{\eta_{0}}\int^{L}_{0}A_{b}(y)e^{-i\Delta \beta_{\mu} y}dy=\sqrt{\eta_{0}}A_{b,0}L\Gamma_{\mu}.
\end{equation}
The output photon rate at mode $\pm\mu$ in the low-gain regime is 
\begin{equation}
\label{eq:QFC_photon_rate}
n_{\pm\mu,\mathrm{out}}=\frac{\kappa_{\pm\mu,\mathrm{e}}\kappa_{\mp\mu}|g_{\mu}|^{2}}{2\bar{\kappa}(\kappa_{\pm\mu}\kappa_{\mp\mu}/4+\delta_{\mu}^{2})},
\end{equation}
where $g_{\mu}=G_{\mu}/\tau_{a}$ is the parametric gain rate at mode $\pm\mu$, and the average loss rate of the pair is defined as $\bar{\kappa}\equiv(\kappa_{\mu}+\kappa_{-\mu})/2$. For modes close to the degenerate mode, the parametric gain rate is approximately the same given that waveguide SPDC bandwidth is on the order of several THz \cite{ahler2026low}, and therefore, the output photon rate is mainly controlled by modal detuning. At a non-zero pump detuning of $\delta_b=D_{2}\mu^{2}$, the SPDC rate is maximized for mode pair $\pm\mu$, and for the normal-dispersion resonators studied here, a GHz-level negative pump detuning is needed for maximizing SPDC rate at higher $\mu$. 

Figure~\ref{Fig4}\textbf{b} shows the normalized QFC spectra at various SH pump detuning calculated using Eq.~\ref{eq:QFC_photon_rate}, and experimentally measured on an optical spectrum analyzer (OSA). With good agreement between the model and experiment, highest QFC power is concentrated along the white parabolic curve for modes satisfying $\delta_b=D_{2}\mu^{2}$. As expected, red-detuning the SH pump moves the bright pairs outward from the degenerate mode, and a detuning of 5~GHz accesses signal and idler modes spanning more than 2.8~THz. The pump detuning therefore acts as a single, continuously tunable knob that selects which signal–idler pairs of the comb are brightest, without any change to the device temperature or heater power. 

To characterize the device as a photon-pair source, we tune the SH near $m=1$ and select the non-degenerate pair (-3,~4) with two narrowband fiber Bragg filters, detecting the signal and idler photons on superconducting nanowire single-photon detectors (SNSPDs), and photon-counting events are time-tagged with a time-correlated single-photon counter (TCSPC). By fitting the cross-correlation function versus time delay, $g^{(2)}(\tau)$, we extract the internal pair-generation rate (PGR), and the details are discussed in Supplementary Information Section S5. Figure~\ref{Fig4}\textbf{c} shows the back-calculated on-chip PGR at the cavity output and the measured coincidence-to-accidental ratio (CAR) as a function of on-chip SH pump power. We derive an internal pair-generation efficiency of $R/P_{b,\mathrm{in}}=2.37\pm0.01$~MHz/$\upmu$W from a linear fit, and after accounting for the measured cavity escape efficiency of $\approx33$\%, the on-chip pair-generation efficiency at the cavity output is estimated to be $R_{\mathrm{out}}/P_{b,\mathrm{in}}=0.26$~MHz/$\upmu$W. The CAR falls inversely with pump power, as expected for a SPDC process when accidental coincidences are dominated by multi-pair emission, reaching $(2.87\pm0.18)\times10^{4}$ at an on-chip PGR of 5~kHz and remains above 100 for an on-chip PGR of up to 1.1~MHz. Using the measured pair-generation efficiency, quality factors, and escape efficiencies, we derive a normalized efficiency of $\eta_{0}=0.57\times10^{4}$~\%W$^{-1}$cm$^{-2}$ in the nonlinear section, which is lower than that obtained in the SHG measurement. This discrepancy may arise from an imperfect spiral termination that does not fully eliminate the residual SH pump, causing reflection and possibly destructive interference. 

We examine the effect of pump detuning on the envelope of photon-pair correlation by inserting a programmable filter (PF) to isolate pairs generated from different signal-idler modes. Figure~\ref{Fig4}\textbf{d} shows the coincidence rate detected in signal-idler mode-pair indices 1–25 at three SH pump detunings. At $\delta_{b}/2\pi=0$, the rate is highest for the innermost pairs and decreases monotonically with increasing pair index, whereas red-detuning the SH pump to -0.5~GHz and -0.9~GHz shifts the maximum to pairs with higher mode numbers, consistent with the SPDC model (orange curve). Deviations of individual mode pairs from the modeled envelope may originate from variations in $\kappa_{\pm\mu}$ and perturbations of $\omega_{a,\pm\mu}$ from the fitted dispersion.

Finally, as shown in Fig.~\ref{Fig4}\textbf{e}, we investigate the joint spectral correlation of the distinct signal and idler modes 1 to 15 at optimal pump detuning. Coincidences appear only along the anti-diagonal of the $15\times15$ frequency space and are more than 2 orders of magnitude brighter than the off-diagonal background (limited by noise introduced from the PF), indicating strong frequency correlation between the spectrally distinct, simultaneously addressable photon-pair channels. This tunable, bright quantum frequency comb with 37~GHz repetition rate unlocks possibilities for high-rate, frequency-multiplexed entanglement generation and quantum-state engineering \cite{reimer2016generation, lu2023frequency}. 
\subsection{Optical parametric oscillation}
\label{sec:OPO}
\begin{figure*}[t!]
    \centering
    \includegraphics[width=\textwidth]{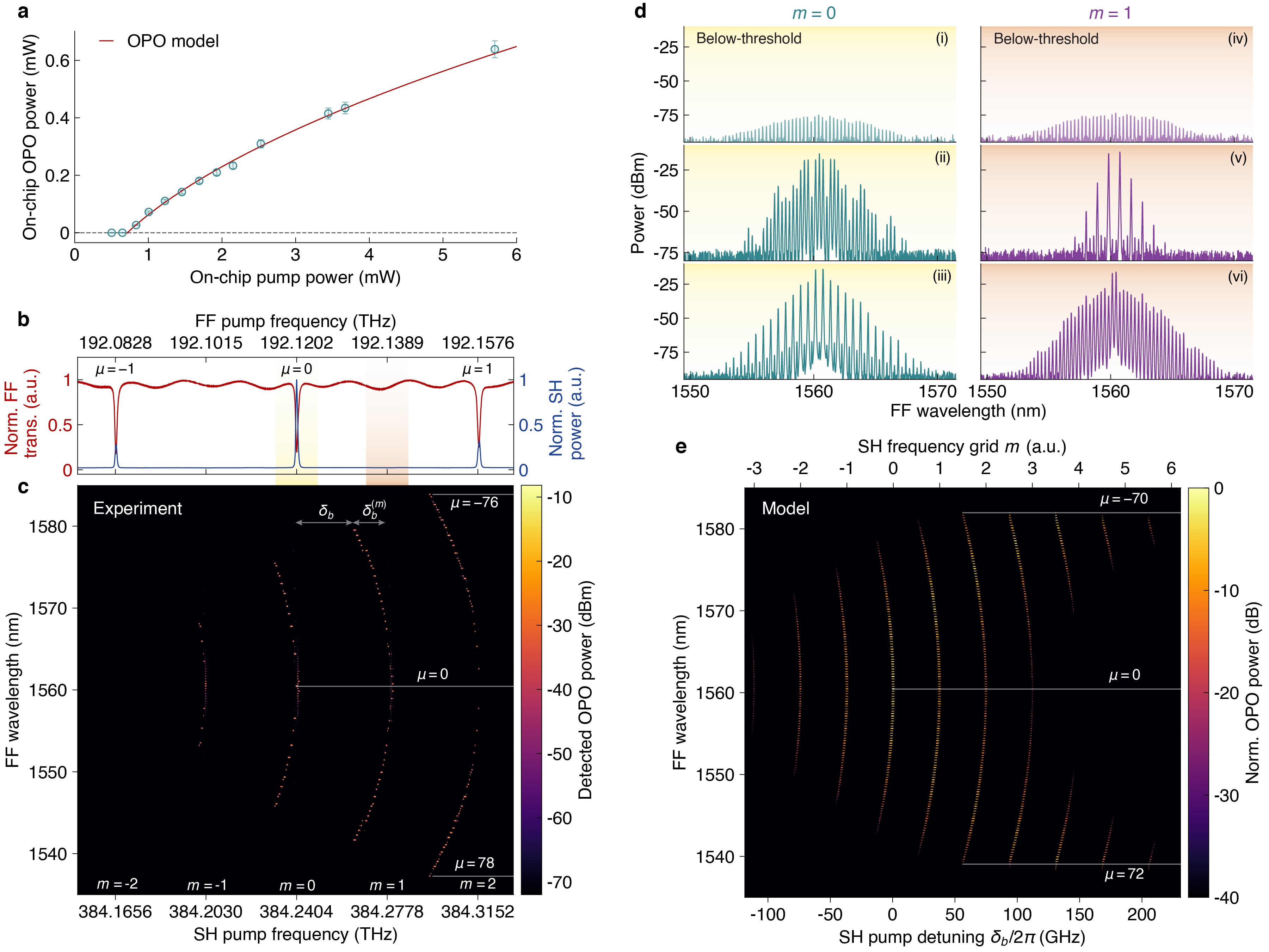}
    \vspace{-15pt}
    \caption{\small \label{Fig5} \textbf{Degenerate and non-degenerate optical parametric oscillation.} \textbf{a}, On-chip OPO power versus on-chip pump power. As pump power is increased, parametric oscillation is observed. By fitting the data to the OPO model (red curve), a threshold power of $P_{\mathrm{th}}=0.71$~mW is obtained. \textbf{b}, Normalized FF transmission (red) and normalized SHG spectrum (blue) when the SH intensity is maximized at the $\mu=0$ mode near 1560.4~nm. \textbf{c}, Tuning between degenerate and non-degenerate OPO when the pump frequency is swept from $m=-2$ to $m=2$. The OPO spectrum is recorded on an optical spectrum analyzer at each pump frequency detuning. The x-scale of \textbf{c} is twice that of \textbf{b}. \textbf{d}, Below-threshold quantum frequency comb and bright OPO comb when: (i)-(iii) the pump is tuned near $m=0$; (iv)-(vi) the pump is tuned at $m=1$. (i) and (iv) are below-threshold, while the other four spectra are above-threshold. \textbf{e}, Calculated OPO tuning using the parametric oscillation model and measured dispersion parameters when the pump frequency is swept from $m=-2$ to $m=6$.
    }
    \vspace{-10pt}
\end{figure*}
When injected SH pump power is high enough to cause sufficiently large parametric gain at FF modes, the down-converted mode becomes unstable, and oscillation occurs when parametric gain exceeds loss \cite{bruch2019chip}. Similar to the SPDC case discussed earlier, the device supports degenerate and non-degenerate OPO operation, with pump detuning being the controlling parameter. For the degenerate mode at zero pump detuning, the oscillation threshold power is given by (see Supplementary Information Section S1):
\begin{equation}
P_{\mathrm{th}}=\frac{\kappa_{a}^{2}\tau_{a}^{2}}{4\eta_{0}L^{2}|\Gamma|^{2}},
\end{equation}
and the above-threshold OPO output power follows
\begin{equation}
\label{eq:OPO_output}
    P_{a,\mathrm{out}}=\frac{\alpha_{b}L\Gamma^{2}}{1-\Gamma}\frac{\kappa_{a,\mathrm{e}}}{\kappa_{a}}P_{\mathrm{th}}\Bigg(\sqrt{\frac{P_{b,\mathrm{in}}}{P_{\mathrm{th}}}} - 1\Bigg).
\end{equation}

With $\eta_{0}$ extracted from the SPDC measurement, $P_{\mathrm{th}}$ for the degenerate mode is estimated to be 0.61~mW. Figure~\ref{Fig5}\textbf{a} shows the on-chip OPO power versus on-chip SH pump power, when the phase-matching condition is optimized by maximizing the SHG power such that SH intensity is maximized at the $\mu=0$ mode near 1560.4~nm, as shown in Fig.~\ref{Fig5}\textbf{b}. Above-background parametric oscillation is detected when the pump power exceeds 0.83~mW. After fitting the data to Eq.~\ref{eq:OPO_output}, we derive an oscillation threshold of $P_{\mathrm{th}}=0.71$~mW, consistent with the estimation from the SPDC measurements. 

At non-zero pump detuning, the OPO emission switches from degenerate to non-degenerate states, where the mode pair with the maximum gain-to-loss ratio has the lowest threshold and therefore oscillates preferentially. Because the SH pump is non-resonant, it can be placed freely on the pump grid, and the oscillating pair ($m-\mu$,~$\mu$) is selected by energy conservation when $\omega_{b}$ is tuned near $\omega_{b}^{(m)}$. When the pump is detuned near pump grid $m=0$, the threshold power for non-degenerate modes $\pm\mu$ is
\begin{equation}
P_{\mu,\mathrm{th}}=\frac{\tau_{a}^{2}}{\eta_{0}L^{2}|\Gamma_{\mu}|^{2}}\bigg[\frac{\kappa_{\mu}\kappa_{-\mu}}{4} + \bigg(\frac{D_{2}}{2}\mu^2-\frac{\delta_{b}}{2}\bigg)^{2}\bigg],
\end{equation}
which is minimized when $\delta_{b}=D_{2}\mu^{2}$, identical to the condition for maximizing the intracavity photon flux in the QFC case. This can be extended to non-zero $m$ by replacing the global pump detuning $\delta_{b}$ with a local detuning relative to the nearest grid point $\delta_{b}^{(m)}$ to account for energy conservation of mode pair ($m-\mu$,~$\mu$).

To study the switching between different OPO modes and their spectral characteristics, we inject 5.7~mW of SH pump and sweep its frequency across $\approx$150~GHz from pump grid $m=-2$ to $m=2$ while recording the OPO spectrum at each pump detuning on the OSA, and the results are shown in Fig.~\ref{Fig5}\textbf{c}. Within each grid interval the oscillating pair begins near the degenerate mode and walks symmetrically outward as the pump is red-detuned, producing a family of bright arcs following $\delta_{b}=D_{2}\mu^{2}$ that repeats once per FSR of the pump grid. Unlike the below-threshold QFC that tunes seamlessly, the above-threshold OPO switches oscillating mode pairs in a step-like manner. This sequential appearance of distinct non-degenerate modes is the $\chi^{(2)}$ analogue of the Eckhaus instability ladder for nonlinear microresonators, in which the bifurcation changes the spatial period and frequency of the waveform \cite{puzyrev2022ladder}. At $m=0$, oscillation extends from the degenerate mode at $\mu=0$ to signal-idler pair (-49,~49) with a separation of 3.7~THz, while the SH pump is red-tuned by 8.6~GHz. The separation between signal and idler modes increases for larger $m$, reaching a maximum of 5.7~THz for mode pair (-76,~78) at $m=2$. 

Noticeably, as seen in Fig.~\ref{Fig5}\textbf{c}, negative local detuning $\delta_{b}^{(m)}$ near a certain grid point favors accessing higher $\mu$ modes, while for large global detuning $\delta_{b}$ across grid points, positive detuning (higher $m$) extends the OPO bandwidth to more widely separated signal-idler modes. This effect is a direct consequence of competition between energy and momentum conservation in the presence of dispersion: with increasingly negative $\delta_{b}^{(m)}$, energy matching pulls the oscillating pair to higher $\mu$, while the phase-mismatch $\Delta \beta_{\mu}$ grows due to normal dispersion at the SH reducing the parametric gain until it can no longer overcome the cavity loss, eventually terminating the oscillation. With large global detuning $\delta_{b}$ to higher $m$, the leading SH dispersion term $\beta_{b}^{(1)}\delta_{b}$ becomes positive and the parametric gain profile set by $\Delta \beta_{\mu}=0$ selects mode pair with larger $\mu$, therefore extending the OPO bandwidth. Taking into account the measured, wavelength-dependent quality factors, the normalized OPO output power calculated using the above-threshold OPO model is shown in Fig.~\ref{Fig5}\textbf{e}. The OPO mode with the highest output power extends from the degenerate mode at $m=0$ to non-degenerate modes with higher $\mu$ as the pump is positively tuned, showing a maximum oscillation bandwidth similar to the experiment. Further increasing the pump detuning above $m=4$ causes the phase-matching optimum to diverge from the energy-matching condition, eventually terminating the parametric oscillations.

At positive local detuning $\delta_{b}^{(m)}$ for $m=-1,0,1$, multiple modes oscillate above threshold to form frequency combs, and the grid index controls whether the degenerate mode participates. Figure~\ref{Fig5}\textbf{d} compares spectra of below-threshold and above-threshold combs for $m=0$, as shown in panels (i)-(iii); and for $m=1$, as shown in panels (iv)-(vi). In both cases, the below-threshold quantum frequency comb (i, iv) and bright OPO comb (ii, iii, v, vi) occupy the same bandwidth of $\approx20$~nm around 1560~nm, with bright comb lines separated by one, two, or four FSRs. This evidence suggests that the above-threshold combs emerge from $\chi^{(2)}$ rather than $\chi^{(3)}$ processes, and comb generation is achievable in both degenerate and non-degenerate OPO configurations. Future work could measure correlations between the below-threshold comb lines to elucidate the quantum formation dynamics of combs in $\chi^{(2)}$ systems \cite{guidry2022quantum}. Frequency comb formation in a CW-pumped, singly resonant $\chi^{(2)}$ cavity has been demonstrated in free space, where a 15-mm-long PPLN enclosed in a bow-tie cavity required 85~mW of SH pump \cite{mosca2018modulation}. The integrated microresonator demonstrated here reaches the comb regime at 3.6~mW (see Supplementary Information S6), more than 20 times lower than the free-space approach. Dissipative quadratic solitons have been predicted for singly resonant degenerate OPOs \cite{nie2020quadratic} and realized in synchronously pumped free-space and hybrid systems \cite{roy2022temporal, englebert2026temporal}, and their generation in an integrated resonator remains an open goal to which the low comb threshold realized here is directly relevant.
\section{Discussion}
\label{sec:discussion}
While the vast majority of cavity-enhanced second-order nonlinear phenomena are studied in doubly resonant structures, our results demonstrate that singly resonant microresonators provide a versatile alternative for second-order nonlinear optics. By resonantly enhancing the FF field while allowing the SH field to propagate without a cavity resonance, this architecture removes the need to align two resonances that are widely separated in wavelength. Despite the absence of SH resonant enhancement, strong nonlinear interactions are preserved owing to the large second-order nonlinearity of InGaP and high intrinsic quality factors of the FF resonances. Overall, these results establish singly resonant InGaP microresonators as a scalable platform that connects efficient frequency conversion, frequency-multiplexed photon-pair generation, and low-threshold parametric oscillation. The combination of compact cavity length, integrated thermo-optic control, and unprecedented nonlinear efficiencies offers a path toward reconfigurable classical and quantum light sources. Additionally, with recent progress in heterogeneously integrated near-visible tunable lasers \cite{castro2025integrated} and the compatibility of InGaP with foundry-ready silicon nitride photonics \cite{thiel2026wafer}, the low-threshold OPO demonstrated here unlocks the possibility of realizing a fully integrated, turn-key OPO source with on-chip pump lasers. Moreover, the broad transparency window of InGaP, from 0.65~$\upmu$m to 11~$\upmu$m \cite{akin2024perspectives}, provides a foundation for extending these capabilities to longer wavelengths, enabling broadband mid-infrared coherent sources that are vital for numerous mid-infrared spectroscopy applications \cite{marandi2012coherence}.

\section{Methods}
\subsection{Device fabrication}
\label{sec:fabrication}
The singly resonant devices are fabricated through heterogeneous integration of thin-film InGaP onto a SiO$_{2}$-on-silicon wafer. 105~nm of In$_{0.49}$Ga$_{0.51}$P thin film is epitaxially grown on top of 500~nm Al$_{0.8}$Ga$_{0.2}$As etch stop layer on a GaAs substrate via metal-organic chemical vapor deposition (MOCVD). Surface passivation of the InGaP surface is performed through ammonium sulfide chemical treatment followed by atomic layer deposition (ALD) of Al$_{2}$O$_{3}$ \cite{li2026high}. The InGaP is then directly bonded to a silicon carrier wafer with a 3-$\upmu$m-thick thermal oxide layer, after O$_{2}$ plasma surface activation. The GaAs substrate and Al$_{0.8}$Ga$_{0.2}$As etch stop are chemically removed with a series of selective wet etching using NH$_{4}$OH:H$_{2}$O$_{2}$ solution and buffered HF solution, followed by another cycle of ammonium sulfide chemical treatment, ALD deposition of Al$_{2}$O$_{3}$, and SiO$_{2}$ hardmask. Photonic structures are defined using deep-ultraviolet (DUV) photolithography with DUV-42P bottom anti-reflection coating (BARC) and UV6-0.8 photoresist. After development, thermal reflow of the photoresist is conducted to reduce its line edge roughness. Waveguide patterns are transferred to the SiO$_{2}$ hardmask by first etching the BARC layer with O$_{2}$ plasma, and then the hardmask using inductively-coupled plasma reactive ion etching (ICP-RIE) with CF$_{4}$/CHF$_{3}$/O$_{2}$ plasma. After hardmask etching, the residual photoresist and organics are removed using multiple cycles of O$_{2}$ plasma exposure and H$_{2}$O$_{2}$ solution immersion \cite{li2026high}. The InGaP waveguide is then etched by ICP-RIE using Cl$_{2}$/H$_{2}$/Ar plasma \cite{parker2011high}, followed by ALD Al$_{2}$O$_{3}$ passivation. Finally, 2~$\upmu$m of SiO$_{2}$ is deposited by plasma-enhanced chemical vapor deposition (PECVD) as top cladding, and integrated thermo-optic heaters are formed by photolithography, electron-beam evaporation of Ti/Pt, and metal liftoff.
\subsection{Measurement}
\label{sec:measurement}
We characterize the fabricated devices in an experimental setup shown in Supplementary Information Section S2. For coupling the FF input and output on the same side of the chip, we use a multi-channel high numerical-aperture fiber array with a mode field diameter of 4~$\upmu$m and a 127~$\upmu$m pitch size (matching the input/output waveguide spacing). The SH light is coupled on the opposite side of the chip using a lensed fiber with a mode field diameter of 2~$\upmu$m. Chip-to-fiber coupling efficiency of the FF and SH light are $\approx50\%$ and $\approx20\%$, respectively, with the exact values depending on the device and fiber array alignment. Since the singly resonant device has only one edge coupler for the SH path, the SH coupling efficiencies are determined from straight waveguides with the same edge coupler designs on the same chip.

For the SHG experiments, we pump the device at the FF with a continuous-wave tunable laser (IR laser, Toptica CTL) followed by an Erbium-doped fiber amplifier (EDFA, Amonics). Output of the SHG is measured on a calibrated power meter (PM, Newport) and a Si photodiode (PD, Thorlabs), while the FF output with the residual pump is monitored on an InGaAs photodiode (PD, Thorlabs) and another PM. For spectral-resolved measurements of the cascaded OPO, part of the FF output is split to an optical spectrum analyzer (OSA, Yokogawa).

For the SPDC/OPO measurements, the SH pump is generated off-chip by frequency-doubling the telecom pump using a commercial periodically poled lithium niobate waveguide module (PPLN, NTT Innovative Devices). Frequency of the SH pump is fine-tuned by adjusting the telecom laser frequency while the SH pump wavelength is referenced on a wavemeter (WM, Bristol Instruments), and the PPLN temperature is simultaneously optimized to achieve maximal SH output power. The OPOs are measured by filtering the FF output with a longpass filter (LPF, Thorlabs) to remove any residual SH pump, and then sent to the OSA and an InGaAs PD. For the SPDC measurements, a programmable filter (PF, Finisar Waveshaper) is used to separate the signal and idler photons. The pairs are subsequently routed to a pair of superconducting nanowire single-photon detectors (SNSPDs, PhotonSpot) and coincidence events are time-tagged with a time-correlated single-photon counter (TCSPC, Swabian Instruments). When using the PF, the aggregate setup loss, including SNSPD efficiency of 85\% and excluding chip-to-fiber coupling, of the signal and idler paths combined is 14.2~dB. The total setup loss is reduced to 11.7~dB when the PF is replaced by a pair of narrow-band fiber Bragg gratings (FBGs, Advanced Optics Solutions) for measuring PGR and CAR of a fixed pair of modes.
\section*{Funding}
\indent  This work was supported by the Defense Advanced Research Projects Agency (Award No. D24AC00166-00) and the NSF (Quantum Foundry Grant No. DMR-1906325 and NRT Training Program Grant No. 2152201). L.T. and L.W. acknowledge support from the NSF Graduate Research Fellowship Program. M.A.G. was supported by an appointment to the Intelligence Community Postdoctoral Research Fellowship Program at the Massachusetts Institute of Technology, administered by Oak Ridge Institute for Science and Education (ORISE) through an interagency agreement between the U.S. Department of Energy and the Office of the Director of National Intelligence (ODNI).
\section*{Acknowledgment}
\indent A portion of this work was performed in the UCSB Nanofabrication Facility, an open access laboratory. We thank Demis John and Noah Dutra for providing useful insight regarding the fabrication processes. 
\section*{Disclosures}
\indent The authors declare no conflicts of interest.
\section*{Data Availability}
\indent The data that support the figures in this paper and other findings of this study are available from the corresponding author on reasonable request.


\thispagestyle{plain}

\def\bibsection{\section*{References}}
\bibliography{references.bib}

\thispagestyle{plain}
\end{document}


\pagestyle{plain}


\title{Supplementary Information\\Singly resonant InGaP microresonators for efficient, low-threshold second-order nonlinear optics}
\author{Yiming Pang}
\email{yimingpang@ucsb.edu}
\affiliation{\protect\hbox{Electrical and Computer Engineering Department, University of California, Santa Barbara, CA 93106, USA}}
\author{Xuefeng Li}
\affiliation{\protect\hbox{Electrical and Computer Engineering Department, University of California, Santa Barbara, CA 93106, USA}}
\author{Lucas Wang}
\affiliation{Physics Department, University of California, Santa Barbara, CA 93106, USA}
\author{Lillian Thiel}
\affiliation{\protect\hbox{Electrical and Computer Engineering Department, University of California, Santa Barbara, CA 93106, USA}}
\author{Melissa A. Guidry}
\affiliation{\protect\hbox{LIGO, Massachusetts Institute of Technology, Cambridge, Massachusetts, USA}}
\author{Joshua E. Castro}
\affiliation{\protect\hbox{Electrical and Computer Engineering Department, University of California, Santa Barbara, CA 93106, USA}}
\author{\\Max Meunier}
\affiliation{\protect\hbox{Electrical and Computer Engineering Department, University of California, Santa Barbara, CA 93106, USA}}
\author{Nergis Mavalvala}
\affiliation{\protect\hbox{LIGO, Massachusetts Institute of Technology, Cambridge, Massachusetts, USA}}
\author{John E. Bowers}
\affiliation{\protect\hbox{Electrical and Computer Engineering Department, University of California, Santa Barbara, CA 93106, USA}}
\affiliation{Materials Department, University of California, Santa Barbara, CA 93106, USA}
\author{Galan Moody}
\email{moody@ucsb.edu}
\affiliation{\protect\hbox{Electrical and Computer Engineering Department, University of California, Santa Barbara, CA 93106, USA}}
\maketitle
\tableofcontents

\newpage
\section*{S1. Model for singly resonant microresonators}
\addcontentsline{toc}{section}{S1. Model for singly resonant microresonators}
\subsection{Waveguide coupled-mode equations}
Here, we consider a singly resonant microresonator that strongly resonates the FF ($a$) mode while being non-resonant to the SH ($b$) mode. The cavity has total length of $L_{\mathrm{c}}$ and a phase-matched nonlinear coupling section with length $L$. For degenerate processes, we can describe the nonlinear interactions in the phase-matched section using the coupled-mode equations for a single-pass waveguide \cite{boyd2008nonlinear}:
\begin{align}
\begin{split}
    \frac{dA_{a}}{dy}&=-i\sqrt{\eta_{0}}A_{a}^{*}A_{b}e^{-i\Delta \beta y} - \frac{\alpha_{a}}{2}A_{a},\\
    \frac{dA_{b}}{dy}&=-i\sqrt{\eta_{0}}A_{a}^{2}e^{i\Delta \beta y} - \frac{\alpha_{b}}{2}A_{b},
\end{split}
\end{align}
where $A_{a}$ and $A_{b}$ represent the normalized electric field amplitudes of the FF and SH waves at $\omega_{a}$ and $\omega_{b}$, respectively. The field amplitude is normalized to the optical power through $P_{i}=|A_{i}|^{2}$. $\alpha_{i}$ is the propagation loss at $\omega_{i}$, and $\Delta \beta=\beta_{b}-2\beta_{a}$ is the phase mismatch between the FF and SH modes. $\eta_{0}$ is the normalized waveguide SHG efficiency in units of W$^{-1}$cm$^{-2}$, and is related to spatial overlap of the propagating modes through \cite{luo2018highly, akin2024ingap}
\begin{equation}
    \eta_{0}=\frac{\omega_{a}^{2}}{2\epsilon_{0}c^{3}n_{a}^{2}n_{b}}\Bigg(\frac{\int_{\chi^{(2)}}\sum_{ijk}\chi_{ijk}^{(2)}E^{*}_{b,i}E_{a,j}E_{a,k}dxdz}{\int_{\text{all}}|\textbf{E}_{a}|^{2}dxdz \sqrt{\int_{\text{all}} |\textbf{E}_{b}|^{2}dxdz}}\Bigg)^{2}.
\end{equation}
Here, $\textbf{E}_{a(b)}$ is the electric field distribution of the FF (SH) mode, $E_{a(b),i}$ is its magnitude in the $i$-direction, and $n_{a(b)}$ its effective refractive index. The integrals $\int_{\chi^{(2)}}dxdz$ and $\int_{\text{all}}dxdz$ denote two-dimensional integration over the cross section of the nonlinear medium and over all space, respectively.

Extending the degenerate case to a non-degenerate one involving a pair of signal ($A_{a,\mu}$) and idler ($A_{a,-\mu}$) modes with $\omega_{a,\mu}\approx\omega_{a,-\mu}$, where $\mu$ is the relative mode number, the coupled-mode equations are:
\begin{align}
\begin{split}
    \frac{dA_{a,\mu}}{dy}&=-i\sqrt{\eta_{0}}A_{a,-\mu}^{*}A_{b}e^{-i\Delta \beta_{\mu} y}- \frac{\alpha_{a}}{2}A_{a,\mu},\\
    \frac{dA_{a,-\mu}}{dy}&=-i\sqrt{\eta_{0}}A_{a,\mu}^{*}A_{b}e^{-i\Delta \beta_{\mu} y}- \frac{\alpha_{a}}{2}A_{a,-\mu},\\
    \frac{dA_{b}}{dy}&=-i2\sqrt{\eta_{0}}A_{a,\mu}A_{a,-\mu}e^{i\Delta \beta_{\mu} y} - \frac{\alpha_{b}}{2}A_{b}.\\
\end{split}
\end{align}
\subsection{Parametric gain}
Now we consider a degenerate spontaneous parametric down conversion (SPDC) process of the FF mode at $\omega_{a}$, driven by a strong SH pump field at $\omega_{b}$. Assuming that the propagation loss of the fundamental wave in the nonlinear waveguide is small $\alpha_{a}L\ll1$ such that it has negligible effect on the effective parametric gain, we can then treat the resonant FF field as a spatially uniform cavity mode coupled to a bath field at the SH frequency, and we make the assumption that this nonlinear coupling is independent of frequency \cite{collett1991two}. In order to describe the system using coupled-mode theory \cite{gardiner1985input}, we convert the position-dependent parametric gain in the waveguide coupled-mode equations to a single-pass gain per round trip.

Defining the position-dependent parametric gain on $A_{a}^{*}$ as
\begin{equation}
    \gamma(y)\equiv\sqrt{\eta_{0}}A_{b}(y)e^{-i\Delta \beta y},
\end{equation}
the single-pass gain through the nonlinear phase-matched section is
\begin{equation}
G=\int^{L}_{0}\gamma(y)dy=\sqrt{\eta_{0}}\int^{L}_{0}A_{b}(y)e^{-i\Delta \beta y}dy.
\end{equation}
Integrating the full SH equation,
\begin{align}
\begin{split}
    A_{b}(y)
    &=e^{-\alpha_{b}y/2}\bigg[A_{b,0}-i\sqrt{\eta_{0}}A_{a}^{2}\int^{y}_{0}e^{(\alpha_{b}/2+i\Delta \beta)y^{\prime}}dy^{\prime}\bigg]\\
    &=e^{-\alpha_{b}y/2}\bigg[A_{b,0}-i\sqrt{\eta_{0}}A_{a}^{2}\frac{e^{(\alpha_{b}/2+i\Delta \beta)y}-1}{\alpha_{b}/2+i\Delta \beta }\bigg].\\
\end{split}
\end{align}
The first term is the propagating and decaying pump, and the second term is the SH generated by back-conversion of the down-converted light. Integrating $A_{b}(y)$ from $0$ to $L$, the parametric gain through the nonlinear section is
\begin{equation}
\label{eq:Gain_degenerate}
        G=\sqrt{\eta_{0}}A_{b,0}L\Gamma-i\frac{\eta_{0}L(1-\Gamma)}{\alpha_{b}/2+i\Delta \beta}A_{a}^{2},
\end{equation}
where we define $\Gamma$ as a dimensionless nonlinear coupling ideality factor,
\begin{equation}
    \Gamma=\frac{1-e^{-(\alpha_{b}/2+i\Delta \beta)L}}{(\alpha_{b}/2+i\Delta \beta)L }.
\end{equation}
If SH loss is neglected ($\alpha_{b}=0$), we recover the well-known $\mathrm{sinc}$ function solution for the phase-matching condition:
\begin{equation}
    \Gamma=\frac{1-e^{-i\Delta \beta L}}{i\Delta \beta L}=\mathrm{sinc}\bigg(\frac{\Delta \beta L}{2}\bigg)e^{-i\Delta \beta L/2}.
\end{equation}
In the expression for the single-pass parametric gain in Eq. \ref{eq:Gain_degenerate}, the first term is gain from the SH pump field, and the second term is gain (or loss, in this case) due to back-conversion of the FF field, which plays an important role in both pump-depleted second-harmonic generation and above-threshold optical parametric generation.

For the non-degenerate case, the single-pass parametric gain for mode $\pm\mu$ is
\begin{equation}
        G_{\mu}=\sqrt{\eta_{0}}\int^{L}_{0}A_{b}(y)e^{-i\Delta \beta_{\mu} y}dy
        =\sqrt{\eta_{0}}A_{b,0}L\Gamma_{\mu}-i\frac{2\eta_{0}L(1-\Gamma_{\mu})}{\alpha_{b}/2+i\Delta \beta_{\mu}}A_{a,\mu}A_{a,-\mu},
\end{equation}
where the phase-mismatch is $\Delta \beta_{\mu}=\beta_{b}-\beta_{a,\mu}-\beta_{a,-\mu}$, and
\begin{equation}
    \Gamma_{\mu}=\frac{1-e^{-(\alpha_{b}/2+i\Delta \beta_{\mu})L}}{(\alpha_{b}/2+i\Delta \beta_{\mu})L }.
\end{equation}
\subsection{Second-harmonic generation}
The effective Hamiltonian for second-harmonic generation and degenerate parametric down-conversion in a singly resonant system is
\begin{equation}
    H = \delta_{a}a^{\dagger}a+g(a^{\dagger})^{2}+g^{\ast}a^2,
\end{equation}
where $a$ denotes the bosonic operator of the degenerate FF mode, and $\delta_{a}=\omega_{a}-\omega_{b}/2$ is the cavity resonance detuning of the FF mode at $\omega_{a}$ relative to the SH field at $\omega_{{b}}$. $g$ is the effective nonlinear coupling rate between the FF and SH photons.

For SHG with a classical FF pump, the dynamics of the system can be described by the Langevin equation \cite{guo2017parametric},
\begin{equation}
\label{eq:SHG_Langevin}
    \frac{da}{dt}=-(i\delta_{a}+\frac{\kappa_{a}}{2})a-i2ga^{\dagger}+i\sqrt{\kappa_{a,\mathrm{e}}}s_{\mathrm{in}},
\end{equation}
where $|a|^{2}$ is the intracavity photon number, $\kappa_{a}$ and $\kappa_{a,\mathrm{e}}$ are the total and extrinsic loss rates of mode $a$ defined by $Q_{a}=\omega_{a}/\kappa_{a}$ and $Q_{a,\mathrm{e}}=\omega_{a}/\kappa_{a,\mathrm{e}}$, respectively. $s_{\mathrm{in}}$ is the driving field amplitude at $\omega_{a}$, and the input FF power in the bus waveguide is $P_{a, \mathrm{in}}=\hbar\omega_{a}|s_{\mathrm{in}}|^{2}$.

Assuming a spatially uniform intracavity field amplitude of $A_{a}=\sqrt{P_{a,\mathrm{cav}}}e^{i\phi_{a}}$ for the FF mode, the SH field amplitude is
\begin{equation}
    A_{b,L}
    =-i\sqrt{\eta_{0}}A_{a}^{2}e^{-\alpha_{b}L/2}\frac{e^{(\alpha_{b}/2+i\Delta \beta)L}-1}{\alpha_{b}/2+i\Delta \beta }.\\
\end{equation}
When perfect phase-matching is achieved, the phase of the generated SH field is locked to the FF pump by $\phi_{b}=2\phi_{a}-\pi/2$. In any case, the SH output power is
\begin{equation}
    P_{b, \mathrm{out}}=|A_{b,L}|^{2}=\eta_{0}P_{a, \mathrm{cav}}^{2}L^{2}|\Gamma|^{2}.
\end{equation}
\subsubsection{Non-depleted approximation}
In the low-power regime, namely $2g\ll\kappa_{a}/2$, back-conversion of the SH to the FF field can be ignored, and the Langevin equation becomes
\begin{equation}
    \frac{da}{dt}=-(i\delta_{a}+\frac{\kappa_{a}}{2})a+i\sqrt{\kappa_{a, \mathrm{e}}}s_{\mathrm{in}},
\end{equation}
with the steady-state solution ($da/dt=0$) of
\begin{equation}
    a_{\mathrm{ss}}=\frac{i\sqrt{\kappa_{a, \mathrm{e}}}}{i\delta_{a}+\kappa_{a}/2}s_{\mathrm{in}}.
\end{equation}
At zero pump detuning ($\delta_{a}=0$), the intracavity pump power is then
\begin{equation}
    P_{a, \mathrm{cav}}=\frac{\hbar\omega_{a}|a_{\mathrm{ss}}|^{2}}{\tau_{a}}=\frac{4\kappa_{a, \mathrm{e}}}{\kappa_{a}^{2}\tau_{a}}P_{a, \mathrm{in}},
\end{equation}
where $\tau_{a}$ is the round-trip time. Substituting the intracavity pump power back into the SH output power equation, we arrive at
\begin{equation}
    P_{b, \mathrm{out}}=\eta_{0}L^{2}|\Gamma|^{2}\frac{16\kappa_{a, \mathrm{e}}^{2}}{\kappa_{a}^{4}\tau_{a}^{2}}P_{a, \mathrm{in}}^{2},
\end{equation}
and the SHG conversion efficiency in the non-depletion regime is
\begin{equation}
\eta_{\mathrm{SHG}}=\frac{P_{b, \mathrm{out}}}{P_{a, \mathrm{in}}^{2}}=\eta_{0}L^{2}|\Gamma|^{2}\frac{16\kappa_{a, \mathrm{e}}^{2}}{\kappa_{a}^{4}\tau_{a}^{2}}.
\end{equation}
\subsubsection{Pump depletion}
At large enough SH output power when $2g\sim\kappa_{a}/2$, the generated SH field undergoes back-conversion and interferes destructively with the FF field, depleting the pump. Therefore, the full Langevin equation in Eq.~\ref{eq:SHG_Langevin} needs to be considered. The single-pass pump depletion of the FF through the nonlinear phase-matched section, $S$, is the second term of the parametric gain equation we derived earlier in Eq.~\ref{eq:Gain_degenerate}:
\begin{equation}
        G
        =-iSA_{a}^{2}, \qquad S=\frac{\eta_{0}L(1-\Gamma)}{\alpha_{b}/2+i\Delta \beta}.
\end{equation}
For the resonant FF mode, we convert this single-pass gain (depletion) into a rate by dividing by the round-trip time $\tau_{a}$, and the intracavity photon number gain rate in Eq.~\ref{eq:SHG_Langevin} can be expressed as:
\begin{equation}
    2g=\frac{G}{\tau_{a}},
\end{equation}
where the factor of 2 comes from the Hamiltonian. Using $A_{a}^{2}=\hbar\omega_{a}a^{2}/\tau_{a}$, the Langevin equation becomes
\begin{equation}
    \frac{da}{dt}=-(i\delta_{a}+\frac{\kappa_{a}}{2})a-\frac{\hbar\omega_{a}S}{\tau_{a}^{2}}a^{2}a^{\dagger}+i\sqrt{\kappa_{a,\mathrm{e}}}s_{\mathrm{in}},
\end{equation}
and the steady-state solution, $a_{\mathrm{ss}}$, can be solved numerically to obtain the SH output power.
\subsection{Degenerate SPDC and OPO}
\subsubsection{Below-threshold SPDC}
Now we consider a degenerate spontaneous parametric down conversion process of the FF mode at $\omega_{a}$, driven by a strong SH pump field at $\omega_{b}$. The Langevin equation including vacuum noise from all channels ($a_{\mathrm{in}}$) is
\begin{equation}
    \frac{da}{dt}=-(i\delta_{a}+\frac{\kappa_{a}}{2})a-i2ga^{\dagger}+i\sqrt{\kappa_{a}}a_{\mathrm{in}}.
\end{equation}

Assuming a non-depleted SH pump field, the single-pass parametric gain through the nonlinear phase-matched section when a pump field of strength $A_{b,0}=\sqrt{P_{b, \mathrm{in}}}e^{i\phi_{b}}$ is injected at the waveguide input is
\begin{equation}
G=\sqrt{\eta_{0}}\int^{L}_{0}A_{b}(y)e^{-(i\Delta \beta) y}dy=\sqrt{\eta_{0}}A_{b,0}L\Gamma.
\end{equation}
and is converted to a gain rate via $2g=G/\tau_{a}$. 

Following the derivations in \cite{guo2017parametric}, the intracavity photon number is 
\begin{equation}
     N_{a,\mathrm{cav}}=\braket{a^{\dagger}(t)a(t)}=\frac{|2g|^{2}}{2[(\kappa_{a}/2)^{2}+\delta_{a}^{2}-|2g|^{2}]},
\end{equation}
and the output photon rate extracted from the cavity is
\begin{equation}
n_{a,\mathrm{out}}=\kappa_{a,\mathrm{e}}N_{a,\mathrm{cav}}=\frac{\kappa_{a,\mathrm{e}}|2g|^{2}}{2[(\kappa_{a}/2)^{2}+\delta_{a}^{2}-|2g|^{2}]}.
\end{equation}
When operating far below the threshold, $|2g|^{2}\ll(\kappa_{a}/2)^{2}+\delta_{a}^{2}$, the gain term in the denominator can be dropped. At zero pump detuning, the total (internal) photon pair generation rate (PGR) as a function of pump power is approximately
\begin{equation}
    R=\frac{\kappa_{a} N_{a,\mathrm{cav}}}{2}
    \approx\frac{\eta_{0}L^{2}|\Gamma|^{2}}{\kappa_{a}\tau_{a}^{2}}P_{b,\mathrm{in}}.
\end{equation}
Given the cavity escape efficiency $\eta_{a,\mathrm{esc}}=\kappa_{a,\mathrm{e}}/\kappa_{a}$, the on-chip pair generation rate at the cavity output is
\begin{equation}
    R_{\mathrm{out}}=\eta_{a,\mathrm{esc}}^{2}R
    \approx\frac{\kappa_{a,\mathrm{e}}^{2}\eta_{0}L^{2}|\Gamma|^{2}}{\kappa_{a}^{3}\tau_{a}^{2}}P_{b,\mathrm{in}}.
\end{equation}
\subsubsection{OPO threshold condition}
At sufficiently large parametric gain, the down-converted mode becomes unstable and oscillates when parametric gain exceeds loss:
\begin{equation}
    |2g|^{2} \geq \bigg(\frac{\kappa_{a}}{2}\bigg)^{2} + \delta_{a}^{2}.
\end{equation}
At zero pump detuning, $\delta_{a}=0$, the threshold condition becomes
\begin{equation}
    |2g_{\mathrm{th}}|^{2} =\bigg(\frac{\kappa_{a}}{2}\bigg)^{2},
\end{equation}
and the oscillation threshold power is
\begin{equation}
P_{\mathrm{th}}=\frac{\kappa_{a}^{2}\tau_{a}^{2}}{4\eta_{0}L^{2}|\Gamma|^{2}}.
\end{equation}
\subsubsection{Above OPO threshold}
Above the oscillation threshold, we solve for the rate equation at steady-state to obtain
\begin{equation}
    a=\frac{-i2ga^{\dagger}}{i\delta_{a}+\kappa_{a}/2}.
\end{equation}
Let $a=|a|e^{i\phi_{a}}$ and $g=|g|e^{i\phi_{b}}$ and set the pump detuning to zero, the magnitude and phase relations between the oscillating mode and the parametric gain are:
\begin{equation}
    |2g|=\frac{\kappa_{a}}{2}, \qquad \phi_{b}-2\phi_{a}=\frac{\pi}{2}.
\end{equation}
Here, the round-trip gain is clamped to the loss rate, which is only true if the pump field is allowed to deplete. Therefore, the back-conversion term in Eq.~\ref{eq:Gain_degenerate} needs to be included. In the limit of small phase-mismatch between the pump and the degenerate mode, $\Delta \beta L\ll1$, the nonlinear coupling ideality factor $\Gamma$ is real and only loss-dependent. 

At $\delta_{a}=0$, substituting Eq.~\ref{eq:Gain_degenerate} into the steady-state gain relation $(\kappa_{a}/2)a=-i2ga^{\dagger}=-i(G/\tau_{a})a^{\dagger}$, we obtain
\begin{equation}
    \frac{\kappa_{a}\tau_{a}}{2}=-i\sqrt{\eta_{0}}L\Gamma|A_{b,0}|e^{i(\phi_{b}-2\phi_{a})}-\frac{\eta_{0}L(1-\Gamma)}{\alpha_{b}/2}|A_{a}^{2}|,
\end{equation}
and the above relation is only physical if the gain term is real and positive, thus locking the phase of the OPO mode to that of the pump: $\phi_{b}-2\phi_{a}=\pi/2$. The intracavity OPO power is
\begin{equation}
    P_{a,\mathrm{cav}}=|A_{a}|^{2}=\frac{\Gamma\alpha_{b}L}{2\sqrt{\eta_{0}}L(1-\Gamma)}|A_{b,0}|-\frac{\kappa_{a}\tau_{a}\alpha_{b}L}{4\eta_{0}L^{2}(1-\Gamma)}.
\end{equation}
Setting $|A_{a}|^{2}=0$ recovers the OPO threshold derived earlier in the non-depletion approximation. Since $\sqrt{P_{b,\mathrm{in}}/P_{\mathrm{th}}}=2\sqrt{\eta_{0}}L\Gamma|A_{b,0}|/(\kappa_{a}\tau_{a})$, we can write
\begin{equation}
    P_{a,\mathrm{cav}}=\frac{\kappa_{a}\tau_{a}\alpha_{b}L}{4\eta_{0}L^{2}(1-\Gamma)}\Bigg(\sqrt{\frac{P_{b,\mathrm{in}}}{P_{\mathrm{th}}}} - 1\Bigg).
\end{equation}
The output OPO power extracted from the cavity $P_{a,\mathrm{out}}=\kappa_{a,\mathrm{e}}\tau_{a}P_{a,\mathrm{cav}}$ can be thus expressed as
\begin{equation}
    P_{a,\mathrm{out}}=\frac{\alpha_{b}L\Gamma^{2}}{1-\Gamma}\frac{\kappa_{a,\mathrm{e}}}{\kappa_{a}}P_{\mathrm{th}}\Bigg(\sqrt{\frac{P_{b,\mathrm{in}}}{P_{\mathrm{th}}}} - 1\Bigg).
\end{equation}
If propagation loss of the SH can be ignored $\alpha_{b}L \ll1$, $\Gamma\approx1$ and $1-\Gamma\approx \alpha_{b}L/4$ and the first term is reduced to a constant of 4, and the output power is only limited by cavity escape efficiency of the FF mode.
\subsection{Non-degenerate SPDC and OPO}
\subsubsection{Dispersion and phase-matching}
The resonant frequencies for FF modes near a central frequency $\omega_{a,0}$ can be written as
\begin{equation}
    \omega_{a,\mu}=\omega_{a,0}+D_{1}\mu+\frac{D_{2}}{2}\mu^{2}+\dots, \qquad \mu\in \mathbb{Z},
\end{equation}
where $\mu$ is the relative mode number, $D_{1}/2\pi$ is the free-spectral range (FSR), and $D_{2}$ is the second-order dispersion parameter.

For the SH pump, we define an equidistant pump frequency grid centered around $\omega_{b,0}=2\omega_{a,0}$,
\begin{equation}
    \omega_{b}^{(m)}=2\omega_{a,0}+D_{1}m, \qquad m\in \mathbb{Z}.
\end{equation}
Cavity-enhanced SPDC happens only if the pump frequency is tuned near $\omega_{b}^{(m)}$ due to energy conservation. When the pump frequency $\omega_{b}$ is near $\omega_{b}^{(m)}$ for a chosen $m$, energy conservation is achieved between the pump and non-degenerate mode pairs at $\omega_{a,\mu}$ and $\omega_{a,m-\mu}$, which we label as mode pairing of ($m-\mu$,~$\mu$). For an even integer $m$, the pump frequency grid satisfies $\omega_{b}^{(m)}=2\omega_{a,m/2}$, where degenerate SPDC simultaneously occurs at the $\omega_{a,m/2}$ mode. In contrast, for an odd integer $m$, only non-degenerate SPDC is allowed since $\omega_{b}^{(m)}\neq2\omega_{a,\mu}$ for any integer $\mu$.

For a global pump detuning $\delta_{b}=\omega_{b}-2\omega_{a,0}$, we define the local pump detuning relative to the nearest grid point $m$ as
\begin{equation}
    \delta_{b}^{(m)} \equiv \omega_{b}-\omega_{b}^{(m)}=\delta_{b}-mD_{1}, \qquad |\delta_b^{(m)}|\leq D_{1}/2.
\end{equation}
In the rotating frame of the pump and its equidistant grid, the modal detuning for mode pair ($m-\mu$,~$\mu$) relative to the pump is
\begin{equation}
    \delta_{\mu}^{(m)}=\frac{D_{2}}{4}[\mu^{2}+(m-\mu)^{2}]-\frac{\delta_{b}^{(m)}}{2}.
\end{equation}

Assume perfect phase-matching condition is achieved between the degenerate mode $\omega_{a,0}$ and a pump frequency of $\omega_{b,0}$, where the phase constants satisfy $\beta_{b,0}=2\beta_{a,0}$. Defining the expansion of the phase constant of non-degenerate FF modes around $\omega_{a,0}$, we have
\begin{equation}
    \beta_{a,\mu}\approx\beta_{a,0}+ \beta_{a}^{(1)}D_{1}\mu+\frac{\beta_{a}^{(2)}}{2}(D_{1}\mu)^{2},
\end{equation}
and the phase constant of the SH mode near $\omega_{b,0}$
\begin{equation}
    \beta_{b}\approx\beta_{b,0}+ \beta_{b}^{(1)}\delta_{b}+\frac{\beta_{b}^{(2)}}{2}\delta_{b}^{2},
\end{equation}
where the dispersion parameters are
\begin{equation}
    \beta_{a}^{(n)}=\frac{\partial^{n}\beta_{a,\mu}}{\partial\omega^{n}}\bigg|_{\omega=\omega_{a,0}},\  \beta_{b}^{(n)}=\frac{\partial^{n}\beta_{b}}{\partial\omega^{n}}\bigg|_{\omega=\omega_{b,0}}.
\end{equation}
The phase constant mismatch can be expressed as
\begin{align}
\begin{split}
    \Delta \beta_{\mu}^{(m)}
    &= \beta_{b}-\beta_{a,\mu}-\beta_{a,m-\mu},\\
    &\approx \beta_{b}^{(1)}\delta_{b}+\frac{\beta_{b}^{(2)}}{2}\delta_{b}^{2}
    - \beta_{a}^{(1)}D_{1}m-\frac{\beta_{a}^{(2)}D_{1}^{2}}{2}[\mu^{2}+(m-\mu)^{2}]\\
\end{split}
\end{align}
Using
\begin{equation}
     D_{1}=\frac{2\pi}{\beta_{a}^{(1)}L_{\mathrm{c}}}, \qquad D_{2}=-\frac{\beta_{a}^{(2)}}{\beta_{a}^{(1)}}D_{1}^{2},
\end{equation}
we find
\begin{equation}
    \Delta \beta_{\mu}^{(m)}
    \approx\beta_{b}^{(1)}\delta_{b}+\frac{\beta_{b}^{(2)}}{2}\delta_{b}^{2}- \frac{2\pi m}{L_{\mathrm{c}}}+\frac{\pi D_{2}}{L_{\mathrm{c}}D_{1}}[\mu^{2}+(m-\mu)^{2}].
\end{equation}

In the derivations below, we focus on the $m=0$ case, where the pump is detuned near the degenerate mode and the non-degenerate mode pairing is ($-\mu$,~$\mu$). Nevertheless, the formalism can be easily generalized to the case of large detuning $m\neq0$ by relabeling the indices of the coupled-mode equations for ($m-\mu$,~$\mu$) pairing and including a non-zero $m$ in the definitions listed above of $\delta_{\mu}^{(m)}$ and $\Delta \beta_{\mu}^{(m)}$. For $m=0$, we have
\begin{align}
\begin{split}
\delta_{\mu} &=\frac{D_{2}}{2}\mu^{2}-\frac{\delta_{b}}{2},\\
\Delta \beta_{\mu}&\approx\beta_{b}^{(1)}\delta_{b}+\frac{\beta_{b}^{(2)}}{2}\delta_{b}^{2}+\frac{2\pi D_{2}}{L_{\mathrm{c}}D_{1}}\mu^{2}.
\end{split}
\end{align}
\subsubsection{Below-threshold SPDC}
The effective Hamiltonian of mode pair ($-\mu$, $\mu$) undergoing non-degenerate SPDC is
\begin{equation}
    H = \delta_{\mu}a_{\mu}^{\dagger}a_{\mu}+\delta_{-\mu}a_{-\mu}^{\dagger}a_{-\mu} + g_{\mu}a_{\mu}^{\dagger}a_{-\mu}^{\dagger}+g_{\mu}^{\ast}a_{\mu}a_{-\mu},
\end{equation}
where $\delta_{\mu}=\delta_{-\mu}$ due to symmetry, $g_{\mu}=G_{\mu}/\tau_{a}$ is the gain rate for mode pair ($-\mu$, $\mu$) and the Langevin equations including vacuum noise from all channels are:
\begin{align}
\begin{split}
\dot a_{\mu}&=-(i\delta_{\mu}+\frac{\kappa_{\mu}}{2})a_{\mu}-ig_{\mu}a_{-\mu}^{\dagger}+i\sqrt{\kappa_{\mu}}a_{\mu,\mathrm{in}},\\
\dot a_{-\mu}&=-(i\delta_{-\mu}+\frac{\kappa_{-\mu}}{2})a_{-\mu}-ig_{\mu}a_{\mu}^{\dagger}+i\sqrt{\kappa_{-\mu}}a_{-\mu,\mathrm{in}},
\end{split}
\end{align}
Solving the Langevin equations at steady-state, the intracavity photon numbers are:
\begin{align}
\begin{split}
N_{\mu,\mathrm{cav}}&=\braket{a_{\mu}^{\dagger}(t)a_{\mu}(t)}=\frac{\kappa_{-\mu}|g_{\mu}|^{2}}{2\bar{\kappa}(\kappa_{\mu}\kappa_{-\mu}/4+\delta_{\mu}^{2}-|g_{\mu}|^{2})},\\
N_{-\mu,\mathrm{cav}}&=\braket{a_{-\mu}^{\dagger}(t)a_{-\mu}(t)}=\frac{\kappa_{\mu}|g_{\mu}|^{2}}{2\bar{\kappa}(\kappa_{\mu}\kappa_{-\mu}/4+\delta_{\mu}^{2}-|g_{\mu}|^{2})},
\end{split}
\end{align}
where we define the average loss rate of the pair as $\bar{\kappa}\equiv(\kappa_{\mu}+\kappa_{-\mu})/2$. When operating far below the threshold, $|g_{\mu}|^{2}\ll\kappa_{\mu}\kappa_{-\mu}/4+\delta_{\mu}^{2}$, the gain term in the denominator can be dropped, and the output photon rates are:
\begin{align}
\begin{split}
n_{\mu,\mathrm{out}}&=\kappa_{\mu,\mathrm{e}}N_{\mu,\mathrm{cav}}=\frac{\kappa_{\mu,\mathrm{e}}\kappa_{-\mu}|g_{\mu}|^{2}}{2\bar{\kappa}(\kappa_{\mu}\kappa_{-\mu}/4+\delta_{\mu}^{2})},\\
n_{-\mu,\mathrm{out}}&=\kappa_{-\mu,\mathrm{e}}N_{-\mu,\mathrm{cav}}=\frac{\kappa_{-\mu,\mathrm{e}}\kappa_{\mu}|g_{\mu}|^{2}}{2\bar{\kappa}(\kappa_{\mu}\kappa_{-\mu}/4+\delta_{\mu}^{2})}.\\
\end{split}
\end{align}
At zero pump detuning, the non-degenerate total (internal) photon pair generation rate as a function of pump power is
\begin{equation}
    R=\kappa_{\mu} N_{\mu,\mathrm{cav}}=\kappa_{-\mu} N_{-\mu,\mathrm{cav}}
    \approx\frac{2\eta_{0}L^{2}|\Gamma_{\mu}|^{2}}{\bar{\kappa}\tau_{a}^{2}}P_{b,\mathrm{in}},
\end{equation}
which is twice of the degenerate SPDC rate, if the average loss rate of the non-degenerate modes is identical to the degenerate mode $\bar{\kappa}=\kappa_{a}$ and phase-matching conditions are equally fulfilled $|\Gamma_{\mu}|=|\Gamma|$. Given the cavity escape efficiencies $\eta_{\mu,\mathrm{esc}}=\kappa_{\mu,\mathrm{e}}/\kappa_{\mu}$, and $\eta_{-\mu,\mathrm{esc}}=\kappa_{-\mu,\mathrm{e}}/\kappa_{-\mu}$, the on-chip pair generation rate at the cavity output is
\begin{equation}
    R_{\mathrm{out}}=\eta_{\mu,\mathrm{esc}}\eta_{-\mu,\mathrm{esc}}R
    \approx\frac{2\kappa_{\mu,\mathrm{e}}\kappa_{-\mu,\mathrm{e}}\eta_{0}L^{2}|\Gamma_{\mu}|^{2}}{\bar{\kappa}\kappa_{\mu}\kappa_{-\mu}\tau_{a}^{2}}P_{b,\mathrm{in}}.
\end{equation}
\subsubsection{OPO threshold condition}
At sufficiently large parametric gain, the down-converted mode becomes unstable and oscillates when parametric gain exceeds loss:
\begin{equation}
    |g_{\mu}|^{2} \geq \frac{\kappa_{\mu}\kappa_{-\mu}}{4} + \delta_{\mu}^{2},
\end{equation}
and in terms of pump detuning $\delta_{b}$,
\begin{equation}
    |g_{\mu}|^{2} \geq \frac{\kappa_{\mu}\kappa_{-\mu}}{4} + \bigg(\frac{D_{2}}{2}\mu^2-\frac{\delta_{b}}{2}\bigg)^{2}.
\end{equation}
The threshold condition for mode $\mu$ is
\begin{equation}
    |g_{\mu,\mathrm{th}}|^{2}=\frac{\kappa_{\mu}\kappa_{-\mu}}{4} + \bigg(\frac{D_{2}}{2}\mu^2-\frac{\delta_{b}}{2}\bigg)^{2},
\end{equation}
and the oscillation threshold power is
\begin{equation}
P_{\mu,\mathrm{th}}=\frac{\tau_{a}^{2}}{\eta_{0}L^{2}|\Gamma_{\mu}|^{2}}\bigg[\frac{\kappa_{\mu}\kappa_{-\mu}}{4} + \bigg(\frac{D_{2}}{2}\mu^2-\frac{\delta_{b}}{2}\bigg)^{2}\bigg].
\end{equation}

At zero pump detuning, the oscillation threshold is lowest for the degenerate mode at $\mu=0$. For any non-degenerate mode pair ($-\mu$,~$\mu$), the threshold is minimized for a non-zero pump detuning $\delta_{b}=D_{2}\mu^{2}$. For the InGaP resonator studied here, its strong normal dispersion $D_{2}=-3.75$~MHz means that a GHz-level negative pump detuning is needed for accessing modes with large $\mu$. 

Now recall the expression for phase-mismatch we derived earlier, given that the second term is much smaller than the first, for a negative $D_{2}$, a positive $\delta_{b}$ is required to minimize phase-mismatch. In other words, at small pump detuning, the phase-matching condition competes with the energy-matching condition, which ultimately limits accessing to large mode number $\mu$ for a fixed pump power. However, when the positive pump detuning is large enough $\delta_{b}\approx D_{1}$ such that $m\geq1$, instead of oscillation at the ($-\mu$, $\mu$) pair of modes, pair ($m-\mu$, $\mu$) will oscillate, and the threshold power becomes
\begin{equation}
P_{\mu,\mathrm{th}}^{(m)}=\frac{\tau_{a}^{2}}{\eta_{0}L^{2}|\Gamma_{\mu}|^{2}}\bigg[\frac{\kappa_{\mu}\kappa_{m-\mu}}{4} + \bigg(\frac{D_{2}}{4}[\mu^{2}+(m-\mu)^{2}]-\frac{\delta_{b}-mD_{1}}{2}\bigg)^{2}\bigg].
\end{equation}
which allows pair with large mode numbers $\mu\gg m$ to oscillate when $m$ is increased.
\subsubsection{Above-threshold OPO}
Above the oscillation threshold, we solve for the rate equations at steady-state to obtain:
\begin{equation}
    a_{\mu}=\frac{-ig_{\mu}a_{-\mu}^{\dagger}}{i\delta_{\mu}+\kappa_{\mu}/2}, \qquad a_{-\mu}=\frac{-ig_{\mu}a_{\mu}^{\dagger}}{i\delta_{\mu}+\kappa_{-\mu}/2},
\end{equation}
at a pump detuning of $\delta_{b}=D_{2}\mu^{2}$, the above-threshold gain is clamped to $|g_{\mu}|^{2}=\kappa_{\mu}\kappa_{-\mu}/4$, this gives:
\begin{equation}
    \frac{|a_{\mu}|^{2}}{|a_{-\mu}|^{2}}=\frac{\kappa_{-\mu}}{\kappa_{\mu}}, \qquad 
    \phi_{b}-\phi_{\mu}-\phi_{-\mu}=\frac{\pi}{2}.
\end{equation}
In the limit of small phase-mismatch, $\Delta \beta_{\mu} L\ll1$, $\Gamma_{\mu}$ is real, and we obtain
\begin{equation}
    \frac{\sqrt{\kappa_{\mu}\kappa_{-\mu}}\tau_{a}}{2}=-i\sqrt{\eta_{0}}L\Gamma_{\mu}|A_{b,0}|e^{i(\phi_{b}-\phi_{\mu}-\phi_{-\mu})}-\frac{2\eta_{0}L(1-\Gamma_{\mu})}{\alpha_{b}/2}|A_{\mu}A_{-\mu}|.
\end{equation}
Implementing the phase-locking condition,
\begin{equation}
    \sqrt{P_{\mu,\mathrm{cav}}P_{-\mu,\mathrm{cav}}}=|A_{\mu}A_{-\mu}|=\frac{\Gamma_{\mu}\alpha_{b}L}{4\sqrt{\eta_{0}}L(1-\Gamma_{\mu})}|A_{b,0}|-\frac{\sqrt{\kappa_{\mu}\kappa_{-\mu}}\tau_{a}\alpha_{b}L}{8\eta_{0}L^{2}(1-\Gamma_{\mu})}.
\end{equation}
Setting $|A_{\mu}A_{-\mu}|=0$ recovers the OPO threshold derived earlier in the non-depletion approximation. We can write
\begin{equation}
    \sqrt{P_{\mu,\mathrm{cav}}P_{-\mu,\mathrm{cav}}}=\frac{\sqrt{\kappa_{\mu}\kappa_{-\mu}}\tau_{a}\alpha_{b}L}{8\eta_{0}L^{2}(1-\Gamma_{\mu})}\Bigg(\sqrt{\frac{P_{b,\mathrm{in}}}{P_{\mu,\mathrm{th}}}} - 1\Bigg).
\end{equation}
The output OPO power extracted from the cavity is $P_{\pm\mu,\mathrm{out}}=\kappa_{\pm \mu,\mathrm{e}}\tau_{a}P_{\pm \mu,\mathrm{cav}}$, using $|A_{\mu}|^{2}\kappa_{\mu}=|A_{-\mu}|^{2}\kappa_{-\mu}$, we obtain
\begin{equation}
    P_{\pm\mu,\mathrm{out}}=\frac{\alpha_{b}L\Gamma_{\mu}^{2}}{2(1-\Gamma_{\mu})}\frac{\kappa_{\pm \mu,\mathrm{e}}}{\kappa_{\pm \mu}}P_{\mu,\mathrm{th}}\Bigg(\sqrt{\frac{P_{b,\mathrm{in}}}{P_{\mu,\mathrm{th}}}} - 1\Bigg).
\end{equation}
\subsection{SHG in the presence of a cascaded OPO}
Consider a pair of non-degenerate OPO modes coupled to the degenerate pump mode through cascaded nonlinearities. The effective Hamiltonian is
\begin{equation}
    H = \delta_{a}a^{\dagger}a + \delta_{\mu}a_{\mu}^{\dagger}a_{\mu}+\delta_{-\mu}a_{-\mu}^{\dagger}a_{-\mu}+ g(a^{\dagger})^{2}+g^{\ast}a^2 + g_{\mu}a_{\mu}^{\dagger}a_{-\mu}^{\dagger}+g_{\mu}^{\ast}a_{\mu}a_{-\mu},
\end{equation}
and the Langevin equations of the interacting fields are:
\begin{align}
\begin{split}
\dot a&=-(i\delta_{a}+\frac{\kappa_{a}}{2})a-i2ga^{\dagger}+i\sqrt{\kappa_{a,\mathrm{e}}}s_{\mathrm{in}},\\
\dot a_{\mu}&=-(i\delta_{\mu}+\frac{\kappa_{\mu}}{2})a_{\mu}-ig_{\mu}a_{-\mu}^{\dagger}+i\sqrt{\kappa_{\mu}}a_{\mu,\mathrm{in}},\\
\dot a_{-\mu}&=-(i\delta_{-\mu}+\frac{\kappa_{-\mu}}{2})a_{-\mu}-ig_{\mu}a_{\mu}^{\dagger}+i\sqrt{\kappa_{-\mu}}a_{-\mu,\mathrm{in}},
\end{split}
\end{align}
where the SH field amplitude is
\begin{equation}
    A_{b}(y)= -i\sqrt{\eta_{0}}\frac{e^{i\Delta \beta y}-e^{-\alpha_{b}y/2}}{\alpha_{b}/2+i\Delta \beta}A_{a}^{2} -
    2i\sqrt{\eta_{0}}\frac{e^{i\Delta \beta_{\mu} y}-e^{-\alpha_{b}y/2}}{\alpha_{b}/2+i\Delta \beta_{\mu} }A_{a,\mu}A_{a,-\mu}.
\end{equation}
and the parametric gain
\begin{equation}
    G=-i\frac{\eta_{0}L(1-\Gamma)}{\alpha_{b}/2+i\Delta \beta}A_{a}^{2}-i\frac{2\eta_{0}L(1-\Gamma_{\mu})}{\alpha_{b}/2+i\Delta \beta_{\mu}}A_{a,\mu}A_{a,-\mu}
    =-iSA_{a}^{2}-i2S_{\mu}A_{a,\mu}A_{a,-\mu}.
\end{equation}
Convert this to a rate, we have
\begin{equation}
    -i2g=-\frac{\hbar\omega_{a}S}{\tau_{a}^{2}}a^{2}-2\frac{\hbar\omega_{a}S_{\mu}}{\tau_{a}^{2}}a_{\mu}a_{-\mu},
\end{equation}
and 
\begin{equation}
    -ig_{\mu}=-\frac{\hbar\omega_{a}S}{\tau_{a}^{2}}a^{2}-2\frac{\hbar\omega_{a}S_{\mu}}{\tau_{a}^{2}}a_{\mu}a_{-\mu}.
\end{equation}
The coupled-mode equations can be solved numerically to obtain the steady-state solutions for the pump mode $a_{\mathrm{ss}}$ and the OPO modes $a_{\mu,\mathrm{ss}}$ and $a_{-\mu,\mathrm{ss}}$. Using the phase-locking relations we derived in the SHG and OPO sections, we can write
\begin{equation}
    \phi_{b}=2\phi_{a}-\frac{\pi}{2}=\phi_{\mu}+\phi_{-\mu}+\frac{\pi}{2}, \qquad \Longrightarrow \qquad 2\phi_{a}=\phi_{\mu}+\phi_{-\mu}+\pi.
\end{equation}
Assuming the OPO mode has small mode number $\mu$, it can be approximated that $\Gamma \approx \Gamma_{\mu}$. The output SH power is 
\begin{equation}
    P_{b, \mathrm{out}}=\eta_{0}L^{2}|\Gamma|^{2}\bigg( \frac{\hbar\omega_{a}|a_{\mathrm{ss}}^{2}+2a_{\mu,\mathrm{ss}}a_{-\mu,\mathrm{ss}}|}{\tau_{a}}\bigg)^{2},
\end{equation}
Implementing the phase-locking condition at an arbitrary pump phase $\phi_{a}$, we can see the SH output power is reduced by destructive interference from the non-degenerate OPO pair that consumes the SHG
\begin{equation}
    P_{b, \mathrm{out}}=\eta_{0}L^{2}|\Gamma|^{2}\bigg[ \frac{\hbar\omega_{a}(|a_{\mathrm{ss}}|^{2}-2|a_{\mu,\mathrm{ss}}a_{-\mu,\mathrm{ss}}|)}{\tau_{a}}\bigg]^{2}.
\end{equation}
\section*{S2. Experimental setup}
\addcontentsline{toc}{section}{S2. Experimental setup}
\begin{figure*}[h!]
    \centering
    \includegraphics[width=\textwidth]{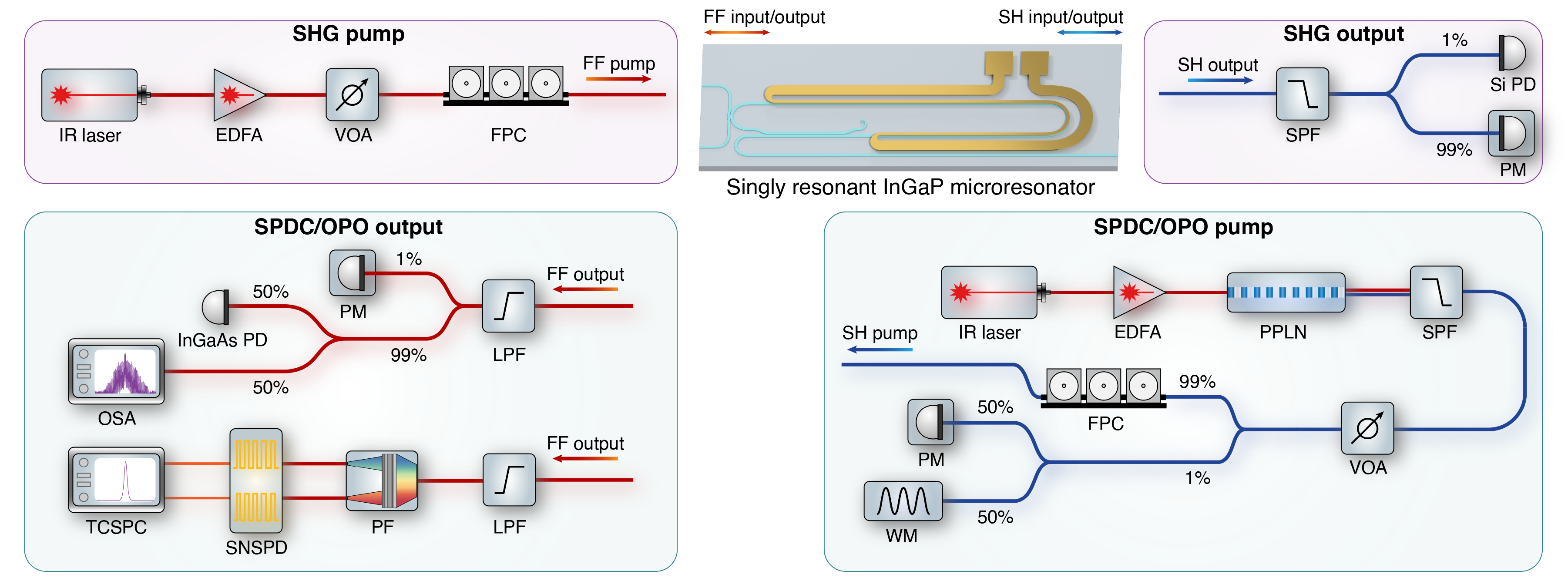}
    \vspace{-15pt}
    \caption{\small \label{FigS1} \textbf{Experimental setup.} A schematic illustration of the experimental setup for the SHG, SPDC, and OPO measurements. IR, infrared (telecom); EDFA, erbium-doped fiber amplifier; VOA, variable optical attenuator; FPC, fiber polarization controller; PM, powermeter; PD, photodiode; SPF, shortpass filter; LPF, longpass filter; PF, programmable filter; PPLN, periodically poled lithium niobate waveguide; OSA, optical spectrum analyzer; WM, wavemeter; SNSPD, superconducting nanowire single-photon detector; TCSPC, time-correlated single-photon counter.
    }
    \vspace{-10pt}
\end{figure*}
\section*{S3. Resonance lineshape fitting}
\addcontentsline{toc}{section}{S3. Resonance lineshape fitting}
In high-Q resonators, roughness-induced Rayleigh backscattering introduces coupling between the forward and backward propagating electric fields, and the resonance lineshape is split into a doublet rather than a Lorentzian singlet \cite{gorodetsky2000rayleigh}. The scattering coupling rate should be a complex coefficient in the form of $\kappa_{\mathrm{s}}=\kappa_{\mathrm{s,r}} + i\kappa_{\mathrm{s,i}}$, as the scattering can lead to both coherent build-up of the counter-propagating field and also dissipative coupling to radiative modes, which results in interference and thus various lineshapes \cite{pfeiffer2018ultra}. The coupling between forward and backward propagating fields can be modeled by:
\begin{equation}
    \begin{split}
        \frac{da_{\mathrm{cw}}}{dt}&=-(i\Delta\omega+\frac{\kappa}{2})a_{\mathrm{cw}}+i\frac{\kappa_{\mathrm{s}}}{2}a_{\mathrm{ccw}}+i\sqrt{\kappa_{\mathrm{e}}}s_{\mathrm{in}},\\
        \frac{da_{\mathrm{ccw}}}{dt}&=-(i\Delta\omega+\frac{\kappa}{2})a_{\mathrm{ccw}}+i\frac{\kappa_{\mathrm{s}}}{2}a_{\mathrm{cw}},
    \end{split}
\end{equation}
where $a_{\mathrm{cw}}$ and $a_{\mathrm{ccw}}$ represent the forward, clockwise propagating wave and the backward, counter-clockwise propagating wave, respectively. $\Delta\omega=\omega_{0}-\omega$ is the cavity resonance detuning, $\kappa=\kappa_{\mathrm{i}} + \kappa_{\mathrm{e}}$ is the total loss rate as a combination of intrinsic and extrinsic (coupling) energy loss rate of the resonator, and $s_{\mathrm{in}}$ is the laser driving field at the waveguide input. Using the steady-state solution and applying the input-output theory, the normalized power transmission of the bus waveguide is
\begin{equation}
\label{eq:transmission}
    T=\bigg\lvert\frac{s_{\mathrm{out}}}{s_{\mathrm{in}}}\bigg\rvert^{2}=\bigg\lvert1-\frac{\kappa_{\mathrm{e}}(i\Delta\omega+\kappa/2)}{(i\Delta\omega+\kappa/2)^{2}+(\kappa_{\mathrm{s}}/2)^{2}}\bigg\rvert^{2}.
\end{equation}

\begin{figure*}[h!]
    \centering
    \includegraphics[width=\textwidth]{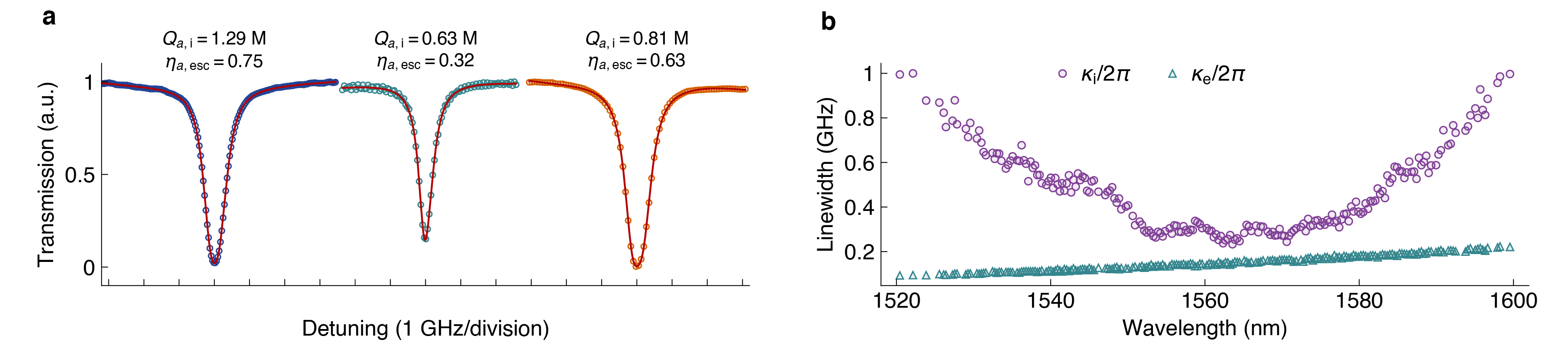}
    \vspace{-15pt}
    \caption{\small \label{FigS2} \textbf{Resonance lineshape fitting and extracted loss rates.} \textbf{a}, Normalized transmission spectra and resonance fits of: the highest intrinsic quality factor resonance in Fig.~2\textbf{e} of the main text (left); the undercoupled resonator used for the SHG/OPO experiments (middle); and the critically coupled resonator used in the SHG experiments (right). The resonance lineshapes on the left ($|\kappa_{\mathrm{s}}|/2\pi\approx0.34$~GHz) and right ($|\kappa_{\mathrm{s}}|/2\pi\approx0.37$~GHz) panels are broadened into doublets as the scattering loss rates being close to the total loss rates ($\kappa/2\pi\approx0.6$~GHz), resulting in close to zero extinctions at over-coupling conditions. \textbf{b}, Intrinsic and extrinsic linewidth of the undercoupled resonator from 1520~nm to 1600~nm. As expected, the extrinsic loss rate exhibits a monotonically increasing trend due to higher coupling rate of the pulley-coupler at longer wavelengths. In contrary, the intrinsic loss rate displays a parabolic trend and is minimized near 1560~nm.
    }
    \vspace{-10pt}
\end{figure*}

Resonance lineshapes of the resonators studied in this work are fitted using Eq.~\ref{eq:transmission}, with fitting parameters $\kappa_{\mathrm{i}}$, $\kappa_{\mathrm{e}}$, $\kappa_{\mathrm{s,r}}$, and $\kappa_{\mathrm{s,i}}$. Additionally, a third-order polynomial background is fitted to account for any Fabry-Pérot reflection in the bus waveguide. Examples of the fitted lineshapes and wavelength dependence of the loss rates are shown in Fig.~\ref{FigS2}.

\section*{S4. SHG tuning with thermo-optic heaters}
\addcontentsline{toc}{section}{S4. SHG tuning with thermo-optic heaters}
\begin{figure*}[h!]
    \centering
    \includegraphics[width=\textwidth]{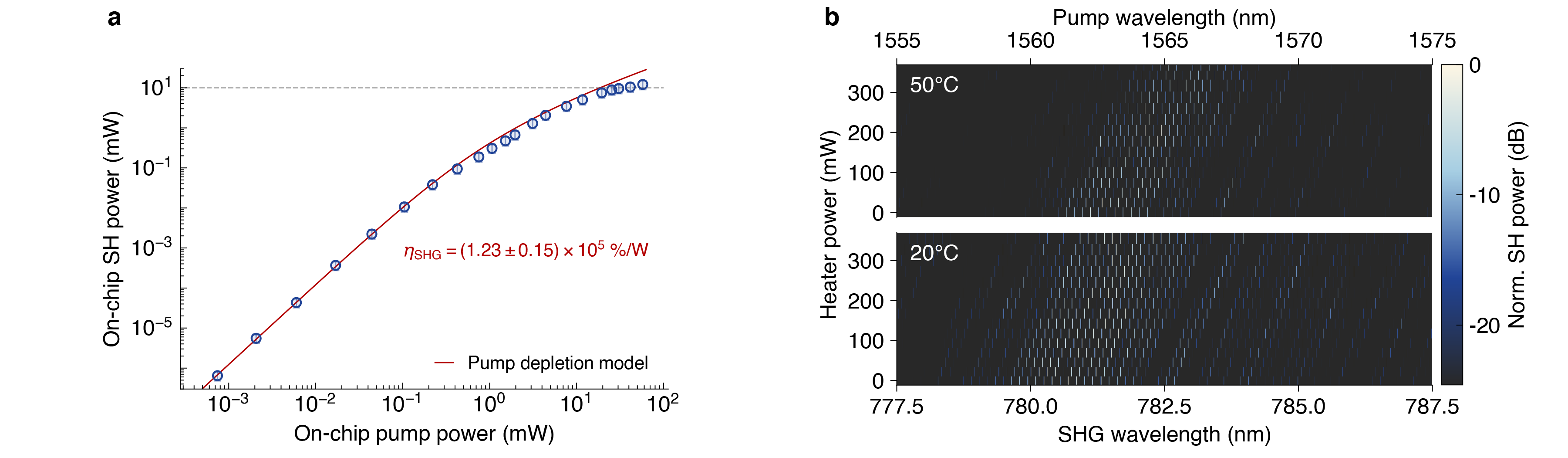}
    \vspace{-15pt}
    \caption{\small \label{FigS3} \textbf{SHG of a critically coupled resonator and thermo-optic tuning.} \textbf{a}, On-chip SH output power versus the on-chip pump power for a critically coupled 2.86~mm long device. Grey dashed line indicates the level of 10~mW SH power. \textbf{b}, Normalized SH power versus wavelength at various integrated heater power, when the chip temperature is set to 20$^{\circ}$C (bottom) and 50$^{\circ}$C (top), respectively.
    }
    \vspace{-10pt}
\end{figure*}
Figure~\ref{FigS3} shows SHG measurements of a critically coupled 2.86~mm long device with $Q_{a,\mathrm{i}}=8.12\times10^{5}$ and $\eta_{a,\mathrm{esc}}=0.63$. The device exhibits high conversion efficiency of $(1.23\pm0.15)\times10^{5}$~\%/W in the non-depletion regime, and generates a maximum of 12.1~mW SH at a pump power of 57.7~mW. By scanning the electrical power applied on the integrated heaters from 0~mW to 360~mW, the SH output wavelength within the 6-dB bandwidth can be tuned freely across 1.8~THz from 779.8~nm to 783.5~nm, while the chip temperature is fixed at 20$^{\circ}$C. Although the maximum amount of electrical power is limited by the current source, further tuning can be achieved by heating up the chip to 50$^{\circ}$C.

\section*{S5. SPDC photon-pair generation and the second-order correlation function}
\addcontentsline{toc}{section}{S5. SPDC photon-pair generation and the second-order correlation function}
For a photon-pair source pumped by a continuous-wave laser source, the second-order correlation function between the signal and idler photons is \cite{signorini2020chip}
\begin{equation}
    g^{(2)}(\tau)=\frac{\braket{a^{\dagger}_{\mathrm{s}}(t+\tau)a^{\dagger}_{\mathrm{i}}(t)a_{\mathrm{i}}(t)a_{\mathrm{s}}(t+\tau)}}{\braket{a^{\dagger}_{\mathrm{s}}(t+\tau)a_{\mathrm{s}}(t+\tau)}\braket{a^{\dagger}_{\mathrm{i}}(t)a_{\mathrm{i}}(t)}}=\frac{R_{\mathrm{si}}(\tau)}{R_{\mathrm{s}}R_{\mathrm{i}}\tau_{\mathrm{b}}},
\end{equation}
where $a^{\dagger}_{\mathrm{s(i)}}$ and $a_{\mathrm{s(i)}}$ are the creation and annihilation operators of photons at the signal (idler) mode, respectively. $R_{\mathrm{si}}(\tau)$ is the signal-idler coincidence event detection rate at time delay $\tau$, $R_{\mathrm{s(i)}}$ is the signal (idler) photon detection rate, and $\tau_{\mathrm{b}}$ is the coincidence time window (bin width). The denominator is the accidental coincidence rate per bin width $R_{\mathrm{acc}}=R_{\mathrm{s}}R_{\mathrm{i}}\tau_{\mathrm{b}}$.

Assuming a Poissonian statistics for the photon counting process, the second-order cross correlation $g^{(2)}(\tau)$ for non-degenerate SPDC can be modeled by the double exponential function. Due to detector jitter and the choice of coincidence time window (bin width), timing uncertainty can be introduced to the coincidence event time-tagging, thus transforming the sharp double exponential distribution into a Gaussian distribution. The $g^{(2)}(\tau)$ can instead be modeled by \cite{guo2017parametric}
\begin{equation}
\label{eq:g2}
    g^{(2)}(\tau)=1+\frac{1}{4R\tau_{\mathrm{c}}}e^{\tau_{\mathrm{w}}^{2}/2\tau_{c}^{2}}[f_{+}(\tau)+f_{-}(\tau)],
\end{equation}
with
\begin{equation}
     f_{\pm}(\tau)=\bigg[1\mp \mathrm{erf}\big(\frac{\tau\pm\tau_{\mathrm{w}}^{2}/\tau_{\mathrm{c}}}{\sqrt{2}\tau_{\mathrm{w}}}\big) \bigg]\cdot e^{\pm\tau/\tau_{\mathrm{c}}},
\end{equation}
where $\tau_{\mathrm{c}}$ is the biphoton coherence time of the SPDC source, $R$ is the internal pair-generation rate (PGR), and $\mathrm{erf}(x)$ is the error function. The total timing uncertainty of the setup is $\tau_{\mathrm{w}}=\sqrt{(\tau_{\mathrm{b}}/2)^{2}+2\tau_{\mathrm{j}}^{2}}$ and is primarily due to a combination of the chosen bin width $\tau_{\mathrm{b}}$ and detector jitter $\tau_{\mathrm{j}}$. 

To quantify the noise of the photon-pair source we use the
coincidence-to-accidental ratio (CAR) evaluated over a coincidence window
$\Delta\tau$:
\begin{equation}
    \mathrm{CAR}(\Delta\tau)=\frac{R_{\mathrm{si}}^{\Delta\tau}-R_{\mathrm{acc}}^{\Delta\tau}}{R_{\mathrm{acc}}^{\Delta\tau}},
    \label{eq:CAR}
\end{equation}
where we denote the raw and accidental coincidence rates within the window as:
\begin{equation}
    R_{\mathrm{si}}^{\Delta\tau}=\frac{1}{\tau_{\mathrm{b}}}
    \int_{-\Delta\tau/2}^{+\Delta\tau/2}R_{\mathrm{si}}(\tau)\,d\tau,
    \qquad
    R_{\mathrm{acc}}^{\Delta\tau}=\frac{\Delta\tau}{\tau_{\mathrm{b}}}R_{\mathrm{acc}}.
    \label{eq:windowed}
\end{equation}
Choosing a single bin, $\Delta\tau=\tau_{\mathrm{b}}$, the expression for CAR reduces
to
\begin{equation}
    \mathrm{CAR}(\tau_{\mathrm{b}})=g^{(2)}(0)-1 .
    \label{eq:CARg2}
\end{equation}

\begin{figure*}[h!]
    \centering
    \includegraphics[width=\textwidth]{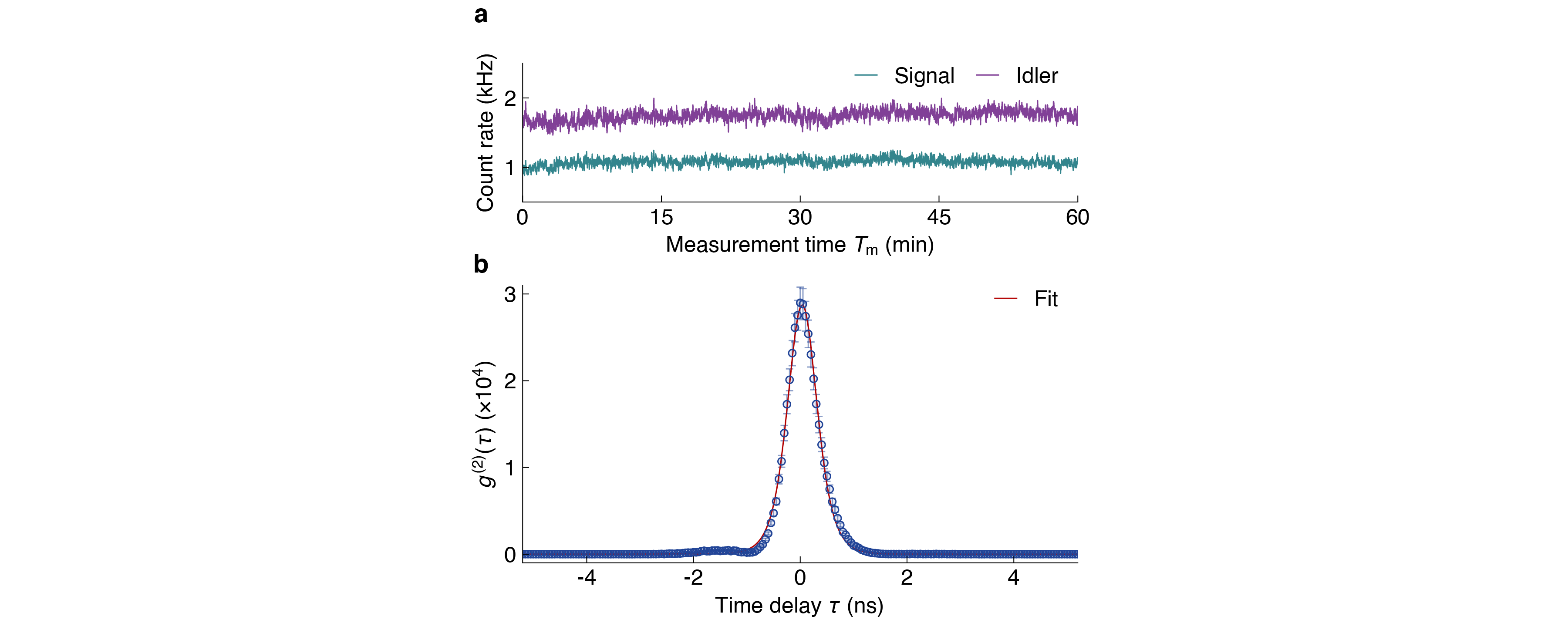}
    \vspace{-15pt}
    \caption{\small \label{FigS4} \textbf{Second-order correlation function.} \textbf{a}, Detected signal and idler count rates as a function of measurement time $T_{\mathrm{m}}$ at a pump power of 18~nW, recorded over an integration time of 1 hour. \textbf{b}, Measured second-order correlation function $g^{(2)}(\tau)$ versus time delay $\tau$. Fit of the $g^{(2)}(\tau)$ function yields a CAR of $(2.87\pm0.18)\times10^{4}$ at an on-chip PGR of 5~kHz.
    }
    \vspace{-10pt}
\end{figure*}

Experimentally, we choose a bin width of $\tau_{\mathrm{b}}=50$~ps and record the
coincidence counts versus time delay, $N_{\mathrm{si}}(\tau)$, over a measurement
time $T_{\mathrm{m}}$ (10~s to 1~hour, depending on the detected rate), which yields
the raw coincidence rate distribution $R_{\mathrm{si}}(\tau)=N_{\mathrm{si}}(\tau)/T_{\mathrm{m}}$. We estimate $R_{\mathrm{acc}}$ directly from the raw data by averaging $R_{\mathrm{si}}(\tau)$ over $M$ bins located far from the correlation peak ($|\tau|>6\tau_{\mathrm{c}}$). A representative $g^{(2)}(\tau)$ histogram fitted with Eq.~\ref{eq:g2}, along with single count rates $R_{\mathrm{s}}$ and $R_{\mathrm{i}}$ monitored during the measurement time $T_{\mathrm{m}}$ are shown in Fig.~\ref{FigS4}. The slight modulation of the $g^{(2)}(\tau)$ lineshape at $\tau\approx-1$~ns is likely due to backscattering of one of the resonator modes.

For error analysis, we assume the photon counting process follows Poissonian statistics, such that the detected coincidence counts has uncertainty of $\sigma_{N}=\sqrt{N}$. The error for PGR ($\sigma_{R}$) is obtained from the covariance matrix of the fitted $g^{(2)}(\tau)$ function. For CAR, its standard error ($\sigma_{\mathrm{CAR}}$) propagated from uncertainty of the raw counts is
\begin{equation}
    \frac{\sigma_{\mathrm{CAR}}}{\mathrm{CAR}}=\sqrt{\frac{1}{g^{(2)}(0)N_{\mathrm{acc}}}+\frac{1}{MN_{\mathrm{acc}}}}.
\end{equation}

\section*{S6. OPO tuning and spectral stability}
\addcontentsline{toc}{section}{S6. OPO tuning and spectral stability}
\begin{figure*}[h!]
    \centering
    \includegraphics[width=\textwidth]{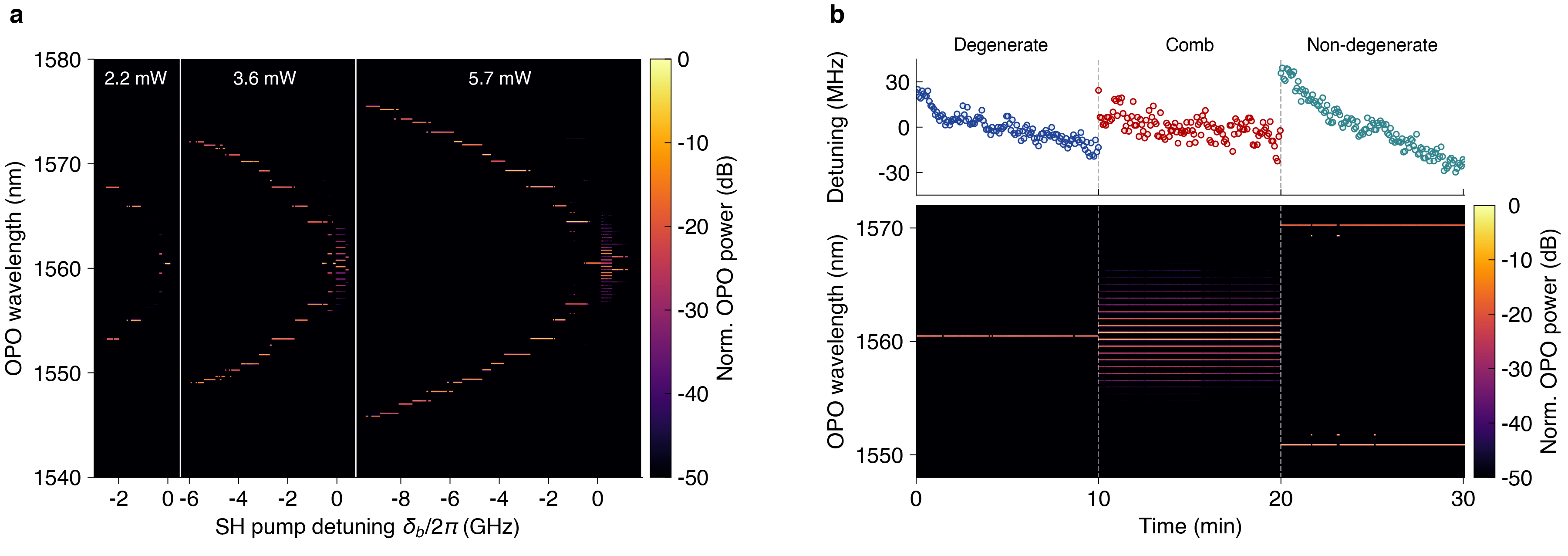}
    \vspace{-15pt}
    \caption{\small \label{FigS5} \textbf{Power-dependence of OPO tuning and OPO spectral stability.} \textbf{a}, OPO spectra versus SH pump detuning $\delta_{b}/2\pi$ when the pump frequency is near grid point $m=0$, at on-chip SH pump power of 2.2~mW (left), 3.6~mW (middle) , and 5.7~mW (right). \textbf{b}, OPO spectral stability in time for the degenerate (left), comb (middle), and non-degenerate OPO states (right), without implementing any active laser locking or coupling stabilization. The top panel shows the pump frequency drift measured on the wavemeter during the spectra acquisition.
    }
    \vspace{-10pt}
\end{figure*}
Figure~\ref{FigS5} \textbf{a} shows the OPO spectra when the SH pump frequency is swept near grid point $m=0$, at three different on-chip SH pump power levels. At 2.2~mW of pump power, only the degenerate mode and 3 pairs non-degenerate OPO modes reach threshold. As the pump power increases to 3.6~mW, the number of non-degenerate OPO modes increases to 10 pairs, and eventually to more than 13 pairs at 5.7~mW (excluding comb lines). Formation of a non-degenerate, 2-FSR frequency comb is observed at 3.6~mW, while 1-FSR comb with the degenerate mode is accessed at 5.7~mW of pump power. 

Additionally, we investigate the spectral stability of different OPO states under a free-running CW pump. As shown in Fig.~\ref{FigS5} \textbf{b}, we show OPO spectra taken over 10 minutes for the degenerate OPO (left), a 2-FSR comb (middle), and a pair of non-degenerate OPO (right) modes by fixing the pump laser frequency and acquiring a spectrum every 5 seconds, while monitoring the pump frequency drift on a wavemeter. Each OPO state remains spectrally stable for the majority of the measurement time, except for a few instances where the degenerate OPO switches to the no-OPO state and the non-degenerate OPO mode-hops to another neighboring pair, but all reverts to the original state after a few seconds. The occasional instability is likely due to fluctuations of fiber-to-chip coupling at the SH pump path, where no active stabilization scheme is used. Although the free-running laser has a constant red-shift of approximately 6~MHz/min due to internal and environmental fluctuations, this small drift in frequency does not seem to have a considerable impact on the OPO stability in the time frame considered here.


\thispagestyle{plain}

\def\bibsection{\section*{References}}
\bibliography{references.bib}

\thispagestyle{plain}